\documentclass[review,12pt]{elsarticle}

\usepackage{hyperref}

\journal{Electrochimica Acta}

\biboptions{sort&compress}
\usepackage{amsmath} 
\usepackage{amssymb}
\usepackage{bm} 
\usepackage{stmaryrd} 
\usepackage{mathtools} 
\usepackage{upgreek}
\usepackage[margin=2cm]{geometry}
\usepackage{changepage} 
\usepackage{soul, xcolor}
\sethlcolor{yellow}

\usepackage[capitalise, nameinlink]{cleveref}
\crefname{appendix}{}{}

\usepackage{subcaption} 
\usepackage{placeins}

\usepackage{setspace}
\begin{document}

\begin{frontmatter}
\title{Corrosion rates under charge-conservation conditions}

\author[1]{Tim Hageman}
\author[2]{Carmen Andrade}
\author[1]{Emilio Martínez-Pañeda \corref{mycorrespondingauthor}}
\cortext[mycorrespondingauthor]{Corresponding author}
\ead{e.martinez-paneda@imperial.ac.uk}
\address[1]{Department of Civil and Environmental Engineering, Imperial College London, London SW7 2AZ, UK}
\address[2]{International Center of Numerical Methods in Engineering (CIMNE), Madrid 28010, Spain}

\begin{abstract}
Laboratory and numerical corrosion experiments impose an electric potential on the metal surface, differing from natural corrosion conditions, where corrosion typically occurs in the absence of external current sources. In this work, we present a new computational model that enables predicting corrosion under charge-conservation conditions. The metal potential, an output of the model, is allowed to change, capturing how the corrosion and cathodic reactions must produce/consume electrons at the same rates, as in natural conditions. Finite element simulations are performed over a large range of concentrations and geometric parameters. The results highlight the notable influence of the charge-conservation assumption and pioneeringly quantify corrosion rates under realistic conditions. They further show: (i) the strong coupling between the corrosion rate and the hydrogen and oxygen evolution reactions, (ii) under which circumstances corrosion pits acidify, and (iii) when corrosion is able to become self-sustained lacking oxygen. \\ 
\end{abstract}



\end{frontmatter}

\section{Introduction}
\label{sec:introduction}
Pitting and other associated forms of localised corrosion are responsible for numerous catastrophic failures across the marine, transport, energy and construction sectors \cite{RILEM2021}. However, despite their importance, corrosion pitting phenomena remain poorly understood and are thus difficult to predict \citep{brown1970concept, galvele1978present, VanderWeeen2014, Kovacevic2023}. 
An area in which predicting pitting corrosion is of critical importance is concrete reinforcement \citep{everett1980deterioration, page1982aspects, andrade1978quantitative}. Concrete is a porous material whose pore solution is very alkaline, a perfect passivator to the steel of the reinforcement. However, this passivity is often disrupted by a lowering of the pH of the pore solution due to carbon dioxide penetration through the pore network, or due to the transport of chloride ions to the steel surface. These mechanisms will result in a loss of passivation, which will induce pitting and localised corrosion \citep{tuutti1982corrosion, alonso1988relation}. A similar process takes place in atmospheric corrosion as the thin water layer present in atmospheric conditions also constitutes a resistive constrain for the corrosion to progress \citep{stratmann1983electrochemical}.

Efforts in understanding the mechanisms underlying pitting corrosion go back to the seminal works by Pourbaix \citep{delahay1950potential, pourbaix1973lectures, pourbaix1974applications, pourbaix1990thermodynamics}, who provided the foundations to rationalise the local acidification detected inside corrosion pits \citep{pickering1972mechanism, turnbull2014corrosion}. Both numerical modelling and laboratory experimentation have been extensively used to investigate the local pit chemical conditions and associated corrosion processes \citep{li2019localized, NGUYEN2021109461, Wang2013, Lin2010, Darowicki2004}. 
However, the practicalities of these methodologies imply that assumptions need to be made regarding the electric current source applied to the metal \citep{Sun2021, Gravano1984}. Models and experimental tests typically consider either an applied potential or an applied current \citep{Lin2020, Wang2021, Li2020, Ansari2019, Vagbharathi2014, Sharland1988, Sarkar2013}. In contrast, most cases of pitting corrosion are not supported by external current sources but by counteracting cathodic reactions, such as the oxygen and hydrogen evolution reactions. While the corrosion process produces electrons, these are consumed by reactions on the same metal involving oxygen and hydrogen, causing a strong coupling between these reaction rates and highlighting the need to understand their interplay and resulting implications (e.g., what is the dependence of corrosion reactions on oxygen availability? Can pits become self-sustained due to role of hydrogen reactions?).

In this work, we present a new computational framework that takes the metal potential as an unknown (degree-of-freedom), allowing to mimic the charge-conservation conditions inherent to the corrosion of isolated metals. Thus, the current generated by the corrosion process is required to be simultaneously consumed by cathodic reactions on the same surface, enabling the study of the interactions between corrosion and the supporting surface reactions. The computational electro-chemical framework is detailed in \cref{sec:gov_eq}, detailing the charge-conservation condition. Then, in \cref{sec:3}, the model is used to investigate corrosion behaviour in a paradigmatic configuration (the pencil electrode test), showing the influence of environmental and geometric parameters on the corrosion rate, and highlighting the implications resulting from the requirement to support corrosion rates by cathodic reactions. The circumstances under which acidic corrosion pits develop are also investigated. Moreover, in \cref{sec:Localisation} we study the interaction between anodic and cathodic reactions, showing that under specific circumstances only a limited external area partakes within this process, and unravelling the role of oxygen in starting and maintaining the corrosion rate. Finally, the findings presented throughout this paper are summarised in \cref{sec:conclusions}.

\section{Modelling framework}
\label{sec:gov_eq}
\begin{figure}
    \centering
    \includegraphics{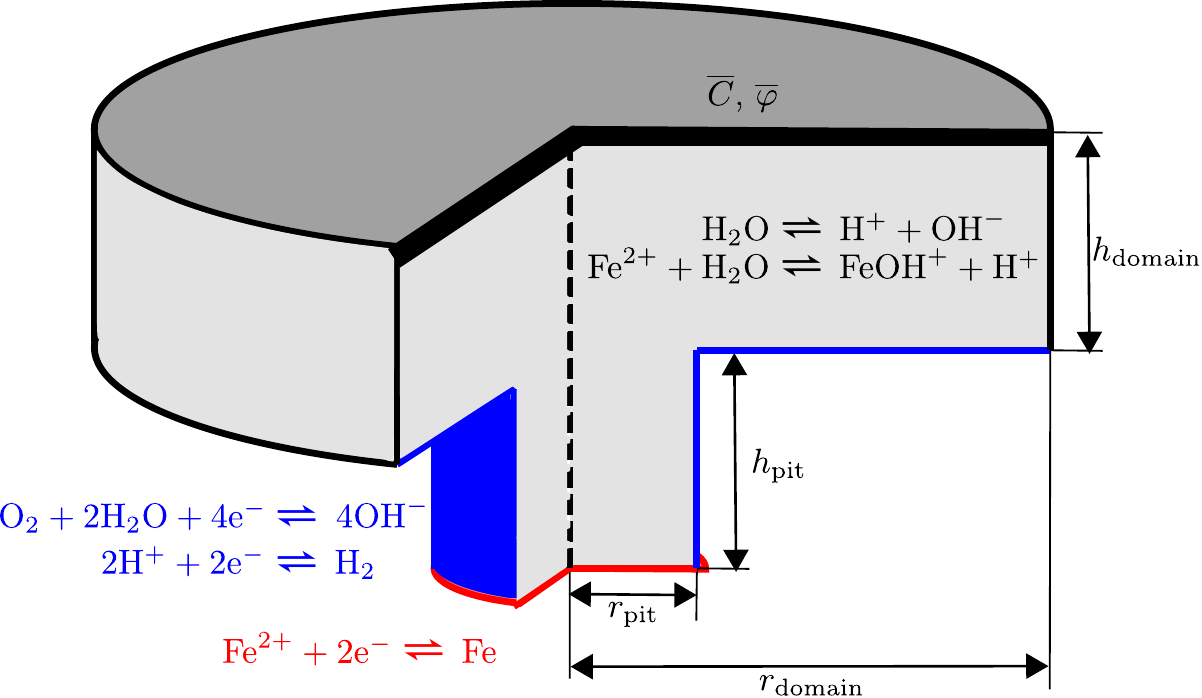}
    \caption{Overview of the simulated domain and reactions occurring therein, with anodic (red) and cathodic (blue) reactions and associated boundaries.}
    \label{fig:domain_pencil}
\end{figure}
We consider the domain shown in \cref{fig:domain_pencil}, which consists of an electrolyte adjacent to a metal surface. This metal surface is divided between anodic and cathodic reacting parts; we assume that corrosion solely happens at the bottom of the pit while cathodic oxygen and hydrogen reactions occur at the sides of the pit and on the exterior surface. The electrolyte itself is described through the electrolyte potential $\varphi$ and the concentrations of ion species $C_\pi$. Specifically, we consider a water-based electrolyte with $\mathrm{H}^+$ and $\mathrm{OH}^-$, corrosion products  $\mathrm{Fe}^{2+}$ and $\mathrm{FeOH}^+$, oxygen $\mathrm{O}_2$, and inert ions $\mathrm{Na}^+$ and $\mathrm{Cl}^-$, to mimic seawater and make the electrolyte more conductive. In addition to these degrees-of-freedom (DOFs), the electric potential of the metal surface $E_m$ is considered as an additional variable, being allowed to vary to balance the rate of ions reacting at the anodic and cathodic surfaces.

While the model described in this section is particularised for corrosion of iron, specifically producing $\mathrm{Fe}^{2+}$, it is more widely applicable to other corrosion products and metals, for instance $\mathrm{Mg}$, $\mathrm{Al}$, or other iron corrosion products. As no assumptions are made about the reactions being activation or diffusion controlled, simulating corrosion through different processes solely requires altering the reaction rate and equilibrium constants. The model itself will, based on these constants, automatically adjust the corrosion rate to match the rate limiting mechanism of the considered metal. And likewise for the cathodic reactions included. While we only consider the oxygen and hydrogen evolution reactions, as these are expected to be the predominant contributions to the cathodic current, additional (electro-) chemical species and reactions can easily be integrated within the presented framework.

\subsection{Electrolyte}
\label{sec:electrolyte}
Within the electrolyte, the conservation and transport of ions are described by the Nernst-Planck equation:
\begin{equation}
    \dot{C}_\pi +\bm{\nabla}\cdot\left(-D_\pi \bm{\nabla}C_\pi\right)+\frac{z_\pi F}{RT}\bm{\nabla}\cdot\left(-D_\pi C_\pi \bm{\nabla}\varphi\right) + R_\pi = 0 \label{eq:massconserv}
\end{equation}
which depends on the diffusivity $D_\pi$, ionic charge $z_\pi$, Faraday constant $F$, gas constant $R$, and reference temperature $T$. In the above equation, the influence of fluid flow is neglected, thus disregarding advective ion transport and solely allowing diffusive transport due to gradients in concentration and electrolyte potential. The reaction term $R_\pi$ is used to include the bulk reactions inside the electrolyte, with the reactions considered being:
\begin{align}
    \mathrm{H}_2\mathrm{O} &\xrightleftharpoons[k_{\mathrm{w}}']{k_{\mathrm{w}}} \mathrm{H}^+ + \mathrm{OH}^-\\
    \mathrm{Fe}^{2+} + \mathrm{H}_2\mathrm{O} &\xrightleftharpoons[k_{\mathrm{fe}}']{k_{\mathrm{fe}}} \mathrm{FeOH}^+ + \mathrm{H}^+
\end{align}
The first of these reactions is the water auto-ionisation reaction, consuming or generating $\mathrm{H}^+$ and $\mathrm{OH}^-$ ions to enforce $\mathrm{pH}+\mathrm{pOH}=14$, as is common in dilute, water-based electrolytes. The second reaction allows the corrosion product to re-generate part of the $\mathrm{H}^+$ ions that are potentially consumed by cathodic reactions. The reaction rates for the water auto-ionisation are given by:
\begin{equation}
    R_{\mathrm{H}^+1}=R_{\mathrm{OH}^-} = k_{\mathrm{w}}C_{\mathrm{H}_2\mathrm{O}} - k_{\mathrm{w}}'C_{\mathrm{H}^+}C_{\mathrm{OH}^-}  = k_{\mathrm{eq}} \left(K_{\mathrm{w}}-C_{\mathrm{H}^+} C_{\mathrm{OH}^-} \right) \label{eq:water_react}
\end{equation}
where the equilibrium constant $K_{\mathrm{w}}$, and a large value for the dummy reaction rate constant $k_{\mathrm{eq}}$ are used to enforce the water auto-ionisation reaction to be in instant equilibrium. For the case of the iron reaction, the relevant reaction rates read:
\begin{align} \label{Re:Iron1}
    R_{\mathrm{Fe}^{2+}}&=-k_{\mathrm{fe}}C_{\mathrm{Fe}^{2+}}+k_{\mathrm{fe}}'C_{\mathrm{H}^+}C_{\mathrm{FeOH}^+} \\
    R_{\mathrm{FeOH}^+}&=R_{\mathrm{H}^+2}=k_{\mathrm{fe}}C_{\mathrm{Fe}^{2+}}-k_{\mathrm{fe}}'C_{\mathrm{H}^+}C_{\mathrm{FeOH}^+}
    \label{Re:Iron2}
\end{align}
The contributions for the hydrogen ions due to these two individual reactions (corrosion and auto-ionisation) are combined together for the reaction rate term in \cref{eq:massconserv}, $R_{\mathrm{H}^+} = R_{\mathrm{H}^+1} + R_{\mathrm{H}^+2}$. Note that, in contrast to reaction (\ref{eq:water_react}), the iron reactions (\cref{Re:Iron1,Re:Iron2}) occur over time scales comparable to the simulated time, and this is characterised by the reaction rates $k_{\mathrm{fe}}$ and $k_{\mathrm{fe}}'$. Although not implicitly included in the model, the eventual equilibrium concentrations reached are related to these two constants by $K_{\mathrm{fe}}=C_{\mathrm{H}^+}C_{\mathrm{FeOH}^+}/C_{\mathrm{Fe}^{2+}}=k_{\mathrm{fe}}/k_{\mathrm{fe}}'$. Furthermore, we focus our attention on dilute solutions and, as a result, no solubility limits or precipitation reactions are included. 

In addition to the Nernst-Planck equations and the reactions above described, the electrolyte potential is obtained by assuming local electroneutrality:
\begin{equation}
    \sum_\pi z_\pi C_\pi = 0 \label{eq:electroneutrality}
\end{equation}
This enforces the conservation of electric current within the electrolyte by requiring the movement of positively charged ions to be locally balanced by the reverse movement of negatively charged ions, such that the total charge remains neutral throughout the electrolyte \citep{Feldberg2000, Sarkar2011, Hageman2022b}. 

\subsection{Surface reactions and charge conservation}
\label{sec:surface_reacts}
The metal surface is assumed to be divided into two distinct parts, separating the anodic corrosion reaction from the cathodic hydrogen and oxygen reactions. This can be viewed as having a local rupture of any protective layers (e.g. due to mechanical damage, dissolution, passivation breakdown) triggering corrosion on this location, while the remainder of the surface retains its corrosion protection. On the anodic part of this metal surface, corrosion is the only reaction considered:
\begin{equation}
    \mathrm{Fe}^{2+}+2\mathrm{e}^-\xrightleftharpoons[k_{\mathrm{c}}']{k_{\mathrm{c}}}\mathrm{Fe}
\end{equation}
whereas the cathodic surface includes oxygen and hydrogen-related reactions:
\begin{align}
    \mathrm{O}_2+2\mathrm{H}_2\mathrm{O} +4\mathrm{e}^-&\xrightleftharpoons[k_{\mathrm{o}}']{k_{\mathrm{o}}} 4 \mathrm{OH}^- \\
    2\mathrm{H}^+ + 2\mathrm{e}^-&\xrightleftharpoons[k_{\mathrm{h}}']{k_{\mathrm{h}}} \mathrm{H}_2 
\end{align}
While both these cathodic reactions are resolved on the complete anodic surface, eliminating the need for prior assumptions, they are not expected to exhibit significant rates at all locations within this surface. As the hydrogen reaction favours acidic environments, it is expected to be the dominant reaction at the sides of the corrosion pit, where a local anodic environment will have been created as a result of the corrosion process and the resulting reactions. In contrast, the oxygen reaction will be prevalent on the exterior surface but negligible on the sides of the pit, where only a limited amount of oxygen will have diffused.

The surface reaction rates (in $\mathrm{mol}/\mathrm{m}^2$) for these three reactions are given through the Butler-Volmer equations as:
\begin{alignat}{2}
    \nu_\mathrm{c} &= k_\mathrm{c} C_{\mathrm{Fe}^{2+}} \exp\left(-\alpha_\mathrm{c} \eta \frac{F}{RT}\right) &&-  k_\mathrm{c}'  \exp\left((1-\alpha_\mathrm{c}) \eta \frac{F}{RT}\right) \label{eq:surf_c}\\
    \nu_\mathrm{o} &= k_\mathrm{o} C_{\mathrm{O}_2} \exp\left(-\alpha_\mathrm{o} \eta \frac{F}{RT}\right) &&- k_\mathrm{o}' C_{\mathrm{OH}^{-}} \exp\left((1-\alpha_\mathrm{o}) \eta \frac{F}{RT}\right) \label{eq:surf_o}\\
    \nu_\mathrm{h} &= k_\mathrm{h} C_{\mathrm{H}^+} \exp\left(-\alpha_\mathrm{h} \eta \frac{F}{RT}\right) && \label{eq:surf_h}  
\end{alignat}
Here, it is assumed that the hydrogen gas produced immediately disappears, allowing us to omit the backward reaction term in Eq. (\ref{eq:surf_h}). The surface reaction rates detailed above can be converted to the experimentally observed reaction currents through the Faraday constant and the number of involved ions: 
\begin{equation}
    i_\mathrm{c} = 2F \nu_\mathrm{c} \qquad i_\mathrm{o} = 4F\nu_\mathrm{o} \qquad i_\mathrm{h} = 2F\nu_\mathrm{h}
\end{equation}
In a similar manner, the reaction constants $k_i$ can be converted to the exchange current densities $i_{0,i}$, separating the forward and backward exchange currents to result in:
\begin{equation}
    i_{0,\mathrm{c}} = 2Fk_\mathrm{c} \qquad i_{0,\mathrm{c}}' = 2Fk_\mathrm{c}' \qquad i_{0,\mathrm{o}} = 4Fk_\mathrm{o} \qquad i_{0,\mathrm{o}}' = 4Fk_\mathrm{o}' \qquad i_{0,\mathrm{h}} = 2Fk_\mathrm{h} 
\end{equation}

The surface reactions depend on the electric overpotential $\eta$, which is related to the jump between the metal and electrolyte potentials through:
\begin{equation}
    \eta = E_\mathrm{m}-\varphi -E_{\mathrm{eq}}
\end{equation}
where $E_\mathrm{m}$ denotes the electric potential of the metal, $\varphi$ the potential of the electrolyte, and $E_{\mathrm{eq}}$ the equilibrium potential (a constant value, dependent on the reaction being considered). The value for the metal potential $E_\mathrm{m}$ is not fixed during the simulations. Instead, it adapts to change the value of the overpotential $\eta$, which in turn accelerates or decelerates the surface reactions to maintain the conservation of electric charge at the surface, see Eqs. (\ref{eq:surf_c})-(\ref{eq:surf_h}). To determine this metal potential, we assume that the metal surface is not attached to any external current source, requiring the charges at the interface to be conserved:
\begin{equation}
    I_\mathrm{c}+I_\mathrm{o}+I_\mathrm{h}=\int_\Gamma i_\mathrm{c}+i_\mathrm{o}+i_\mathrm{h}\;\mathrm{d}\Gamma = \int_\Gamma 2F\nu_\mathrm{c} + 4F\nu_\mathrm{o} + 2F\nu_\mathrm{h} \;\mathrm{d}\Gamma= 0
    \label{eq:current_conservation}
\end{equation}
This assumes that the metal conducts currents well enough to instantly transfer the charge created by the corrosion reaction to anywhere on the cathodic reacting surface without any drop in metal potential. As a result, ions produced by the corrosion reaction need to be instantly removed by the hydrogen and oxygen reactions. This requires the electric potential of the metal to adapt, decreasing the potential to slow down the corrosion reaction while accelerating the cathodic reactions. Physically, this represents iron corroding in the absence of other current sources, such as other nearby metals, and isolated from any external voltage, for instance, due to solely being in contact with the electrolyte and insulators such as concrete.

\subsection{Implementation details}
\label{sec:implementation}
The Nernst-Planck conservation equation, \cref{eq:massconserv}, and the electroneutrality condition, \cref{eq:electroneutrality}, are solved through a standard finite element scheme by converting them to their weak forms, and discretising the concentrations and electrolyte potentials using finite elements. These discretised concentrations and electrolyte potential are also used to obtain the local reaction rates required within \cref{eq:current_conservation}, where the integral over the reacting boundaries is performed through a numerical integration scheme. These three equations are resolved together within a single monolithic scheme. This scheme uses the concentration fields of all species $C_{\pi}$ together with the electrolyte potential field $\varphi$ and the metal potential $E_\mathrm{m}$ (a single scalar) as degrees of freedom, with the discretised equations given in \cref{app:disc}. For the spatial discretisation, axial symmetry is used to translate the three-dimensional problem to a two-dimensional domain, which is then discretised using quadratic triangular finite elements. The temporal discretisation uses a backward Euler scheme. Finally, to provide a stable and oscillation-free solution, a lumped integration scheme is used for all reaction terms \citep{Schellekens1993, Hageman2020a, Hageman2023}. The use of a lumped integration scheme is essential to deliver predictions over technologically-relevant scales. 

One notable assumption made by including the size of the corrosion pit within the geometry of the mesh is that the corrosion process does not dissolve a significant volume of metal over the simulated time. While this could be allowed through, for instance, remeshing \citep{Laycock2001, Scheiner2007, Xiao2011, Sarkar2012}, the extended finite element method \citep{Vagbharathi2014, Duddu2016},  or using a phase-field formulation \citep{Tsuyuki2018, Lin2020, Cui2021}, we choose to focus solely on the effects of the initial geometry and environment. By not having the metal actively dissolving and changing, steady-state results can be obtained based on the geometry of the pit. 

\section{Pencil electrode in charge-conservation conditions}
\label{sec:3}
\begin{table}
    \centering
    \vspace*{-0.5cm}
    \hspace{-1.25cm}
    \begin{tabular}{|c l || r c|}
    \hline
    \hspace{1.0cm}Parameter \hspace{-1.0cm} & & Magnitude \hspace{-1.5cm} & \\
    \hline 
    \hline
         Diffusivity \citep{Lvov2015} & $D_{\mathrm{H}^+}$ & $9.3\cdot10^{-9}$ & $\mathrm{m}^2/\mathrm{s}$  \\
         & $D_{\mathrm{OH}^-}$ & $5.3\cdot10^{-9}$ & $\mathrm{m}^2/\mathrm{s}$  \\
         & $D_{\mathrm{Na}^+}$ & $1.3\cdot10^{-9}$ & $\mathrm{m}^2/\mathrm{s}$  \\
         & $D_{\mathrm{Cl}^-}$ & $2\cdot10^{-9}$ & $\mathrm{m}^2/\mathrm{s}$  \\
         & $D_{\mathrm{Fe}^{2+}}$ & $1.4\cdot10^{-9}$ & $\mathrm{m}^2/\mathrm{s}$  \\
         & $D_{\mathrm{FeOH}^+}$ & $1\cdot10^{-9}$ & $\mathrm{m}^2/\mathrm{s}$  \\
         & $D_{\mathrm{O}_2}$ & $1\cdot10^{-9}$ & $\mathrm{m}^2/\mathrm{s}$ \\
         \hline
         Temperature & $T$ & $293.15$ & $\mathrm{K}$ \\
         \hline
         Water auto-ionisation & $k_{\mathrm{eq}}$ & $10^6$ & $\mathrm{m}^3/\mathrm{mol}/\mathrm{s}$ \\
         constant & $K_{\mathrm{w}}$ & $10^{-8}$ & $\mathrm{mol}^2/\mathrm{m}^6$ \\
         \hline
         Iron reaction constants & $k_{\mathrm{fe}}$ & $0.1$ & $1/\mathrm{s}$ \\
         & $k_{\mathrm{fe}}'$ & $10^{-3}$ & $\mathrm{m}^3/\mathrm{mol}/\mathrm{s}$ \\
         \hline
    \end{tabular}
        \begin{tabular}{|c c || c c|}
    \hline
    \hspace{0.7cm}Parameter\hspace{-0.7cm} & & Magnitude \citep{Holze2007} \hspace{-1.5cm} &  \\
    \hline 
    \hline
         Corrosion & $k_{\mathrm{c}}$ & $0.1/F$ & $\mathrm{mol}/\mathrm{m}^2/\mathrm{s}$  \\
         & $k_\mathrm{c}'$ & $0.1/F$ & $\mathrm{m}/\mathrm{s}$  \\
         & $E_{\mathrm{eq},\mathrm{c}}$ & $-0.4$ & $\mathrm{V}_{\mathrm{SHE}}$  \\
         & $\alpha_{\mathrm{c}}$ & $0.5$ &   \\
         \hline
         Oxygen & $k_{\mathrm{o}}$ & $10^{-4}/F$ & $\mathrm{mol}/\mathrm{m}^2/\mathrm{s}$  \\
         & $k_\mathrm{o}'$ & $10^{-6}/F$ & $\mathrm{m}/\mathrm{s}$  \\
         & $E_{\mathrm{eq},\mathrm{o}}$ & $0.4$ & $\mathrm{V}_{\mathrm{SHE}}$  \\
         & $\alpha_{\mathrm{o}}$ & $0.5$ &   \\
         \hline
         Hydrogen & $k_{\mathrm{h}}$ & $10^{-6}/F$ & $\mathrm{mol}/\mathrm{m}^2/\mathrm{s}$  \\
         & $k_\mathrm{h}'$ & $0$ & $\mathrm{m}/\mathrm{s}$  \\
         & $E_{\mathrm{eq},\mathrm{h}}$ & $0$ & $\mathrm{V}_{\mathrm{SHE}}$  \\
         & $\alpha_{\mathrm{h}}$ & $0.5$ &   \\
         \hline
    \end{tabular} \hspace{-1cm}\\
    \caption{Properties and reaction constants adopted for all simulated cases. Note that the values of the surface reaction constants are scaled by the Faraday constant $F$.}
    \label{tab:properites}
\end{table}
\begin{table}
    \centering
    \hspace*{-1cm}
    \begin{tabular}{| c || c | c | c | c | c | c | c |}
        \hline
         Case &  $C_{\mathrm{O_2}}\;[\mathrm{mol}/\mathrm{m}^3]$ & $C_{\mathrm{Cl}^-}\;[\mathrm{mol}/\mathrm{m}^3]$ & $\mathrm{pH}$& $r_{\mathrm{pit}}\;[\mathrm{cm}]$ & $h_{\mathrm{pit}}\;[\mathrm{cm}]$ & $r_{\mathrm{domain}}-r_{\mathrm{pit}}\;[\mathrm{cm}]$ & $h_{\mathrm{domain}}\;[\mathrm{cm}]$\\ \hline \hline
         A & $\mathbf{10^{-3}-10^3}$ & $600$ & $12$ & $0.5$ & $2$ & $5$ & $\mathbf{0.5-10}$\\ \hline
         B & $1$ & $600$ & $12$ & $0.5$ & $\mathbf{0.5-10}$ & $5$ & $\mathbf{0.5-10}$\\ \hline
         C & $1$ & $600$ & $12$ & $\mathbf{10^{-3}-10}$ & $\mathbf{10^{-3}-10}$ & $5$ & $2$\\ \hline
         D & $1$ & $600$ & $12$ & $\mathbf{0.25-2.5}$ & $2$ & $\mathbf{0.5-10}$ & $2$\\ \hline
         E & $\mathbf{10^{-3}-10^3}$ & $600$ & $\mathbf{1-14}$ & $0.5$ & $2$ & $5$ & $2$\\ \hline
         F & $1$ & $\mathbf{10^{-1}-10^4}$ & $12$ & $0.5$ & $20$ & $5$ & $2$\\ \hline
         Sec. 4 & $0.25$ & $\mathbf{10^{1}-10^3}$ & $12$ & $0.5$ & $2$ & $100$ & $0.5$\\ \hline
    \end{tabular}
    \caption{Initial conditions and geometric parameters for each case study of those considered in \cref{sec:3} and also for the simulations in \cref{sec:Localisation} (last row). The parameters being varied are shown in bold.}
    \label{tab:parameterSweeps}
\end{table}
We consider a typical pencil electrode test geometry, using the domain shown in \cref{fig:domain_pencil}. On the top of this domain, constant $\mathrm{O}_2$, $\mathrm{H}^+$, and $\mathrm{Cl}^-$ concentrations are imposed equal to those detailed in \cref{tab:parameterSweeps}, whereby zero metal ion concentrations are considered ($C_{\mathrm{Fe}^{2+}}=C_{\mathrm{FeOH}^+}=0$). The concentrations of the remaining species at this boundary follow from equilibrium of the auto-ionisation reaction and the electroneutrality condition. Additionally, a reference potential of $\varphi=0\;\mathrm{V}_{\mathrm{SHE}}$ is prescribed at the top of the domain. However, it should be emphasised that this choice of reference potential does not alter the simulation results, but will merely provide an offset to all reported electric potentials when chosen differently. Importantly, no initial metal potential needs to be imposed; it is an output of the model that will instantaneously adapt to fulfil the conservation of charge equation. The geometry of this domain is varied between case studies to investigate the relations between the pit geometry, the external supporting surface, and the corrosion rate. \cref{tab:parameterSweeps} provides the dimensions of geometries considered in each case study. The properties and reaction constants given in \cref{tab:properites} are used in all the simulations. The domain is discretised using quadratic triangular elements with characteristic lengths varying from approximately 2 mm far from the metal surface to 0.2 mm or smaller in the regions near the electrolyte-metal interface. Corrosion behaviour predictions are obtained for a time of 7 days, using an initial time increment of 300 s, which increases by $5\%$ in each time step, up to time increments of $1\;\mathrm{hour}$. Typical calculation times go from 30 minutes to 2 hours for most cases and up to 12 hours for the largest  geometries. These relatively short calculation times are the result of the robustness and efficiency gains associated with the use of lumped integration schemes \citep{Hageman2023}.

\begin{figure}
    \centering
    \includegraphics[width=16cm]{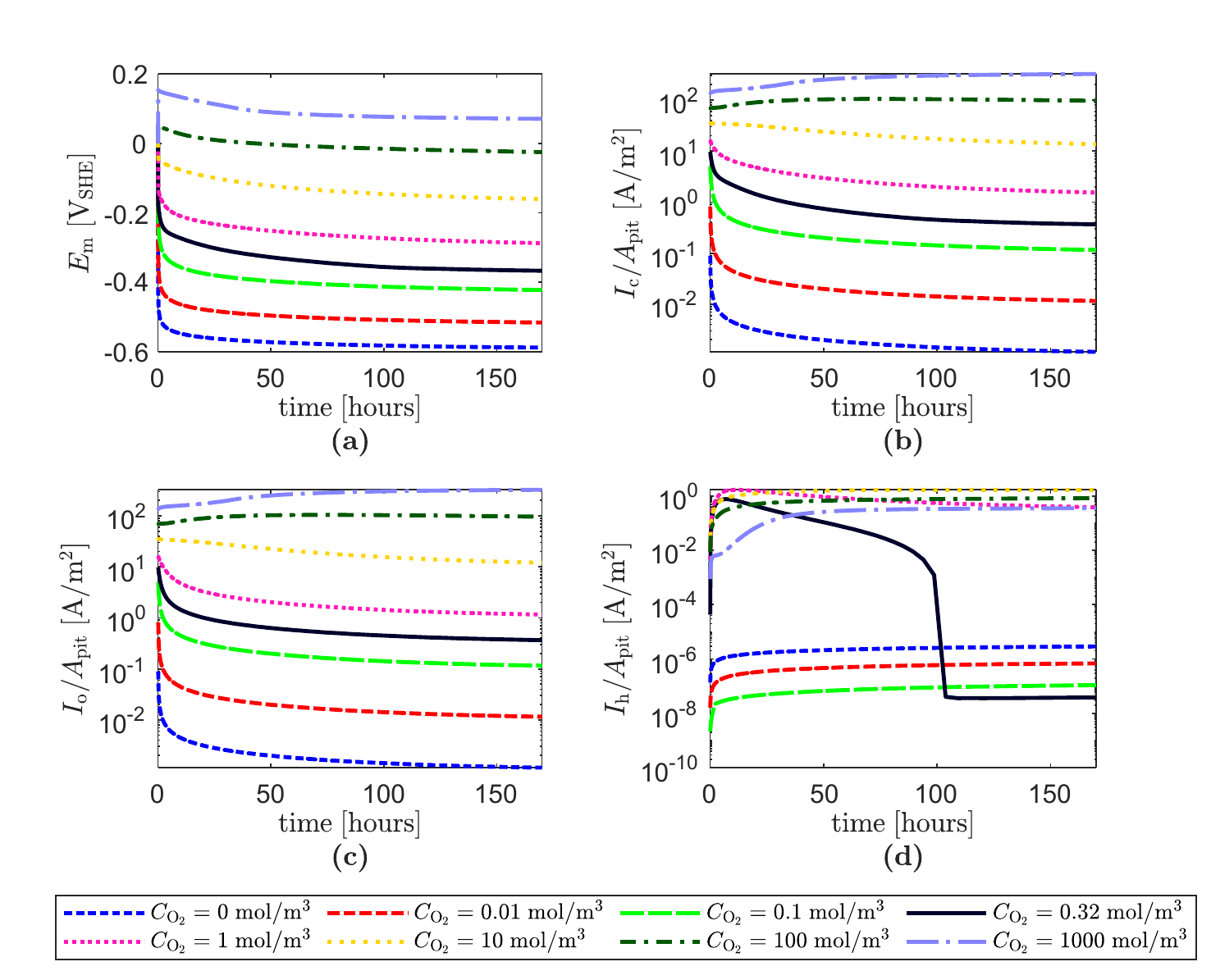}\\
    \caption{Effect of oxygen concentration on the metal potential (a), and corrosion (b) and oxygen (c) and hydrogen (d) reaction currents over the full simulation duration for case study A using a domain height of $\mathrm{h}_{\mathrm{domain}}=5\;\mathrm{cm}$.}
    \label{fig:case1_EM_Current}
\end{figure}

\begin{figure}
    \centering
    \begin{subfigure}{8cm}
         \centering
         \includegraphics[width=8cm]{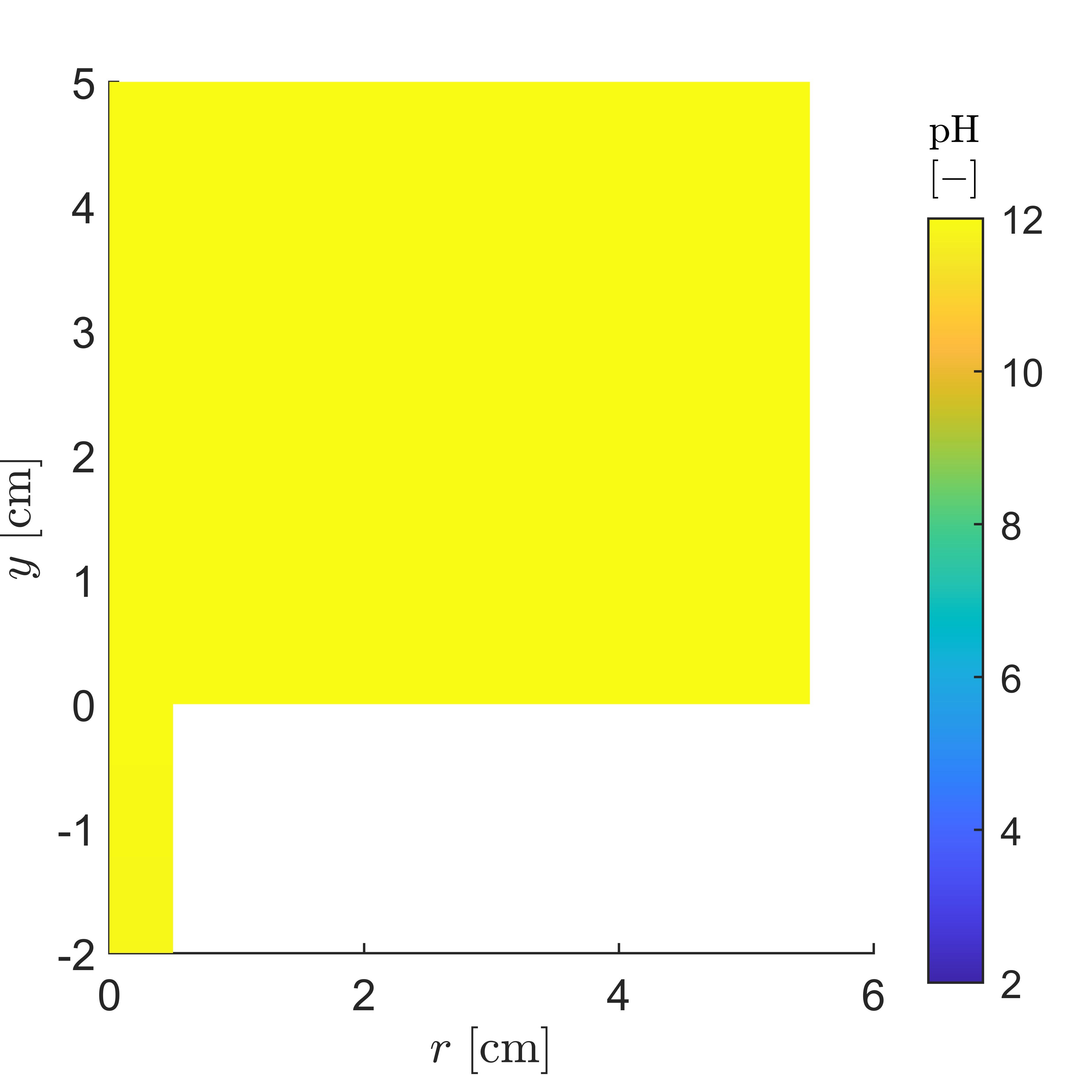}
         \caption{$C_{\mathrm{O}_2}=0.1\;\mathrm{mol}/\mathrm{m}^3$}
         \label{fig:case1_pH_end_5}
    \end{subfigure}
    \begin{subfigure}{8cm}
         \centering
         \includegraphics[width=8cm]{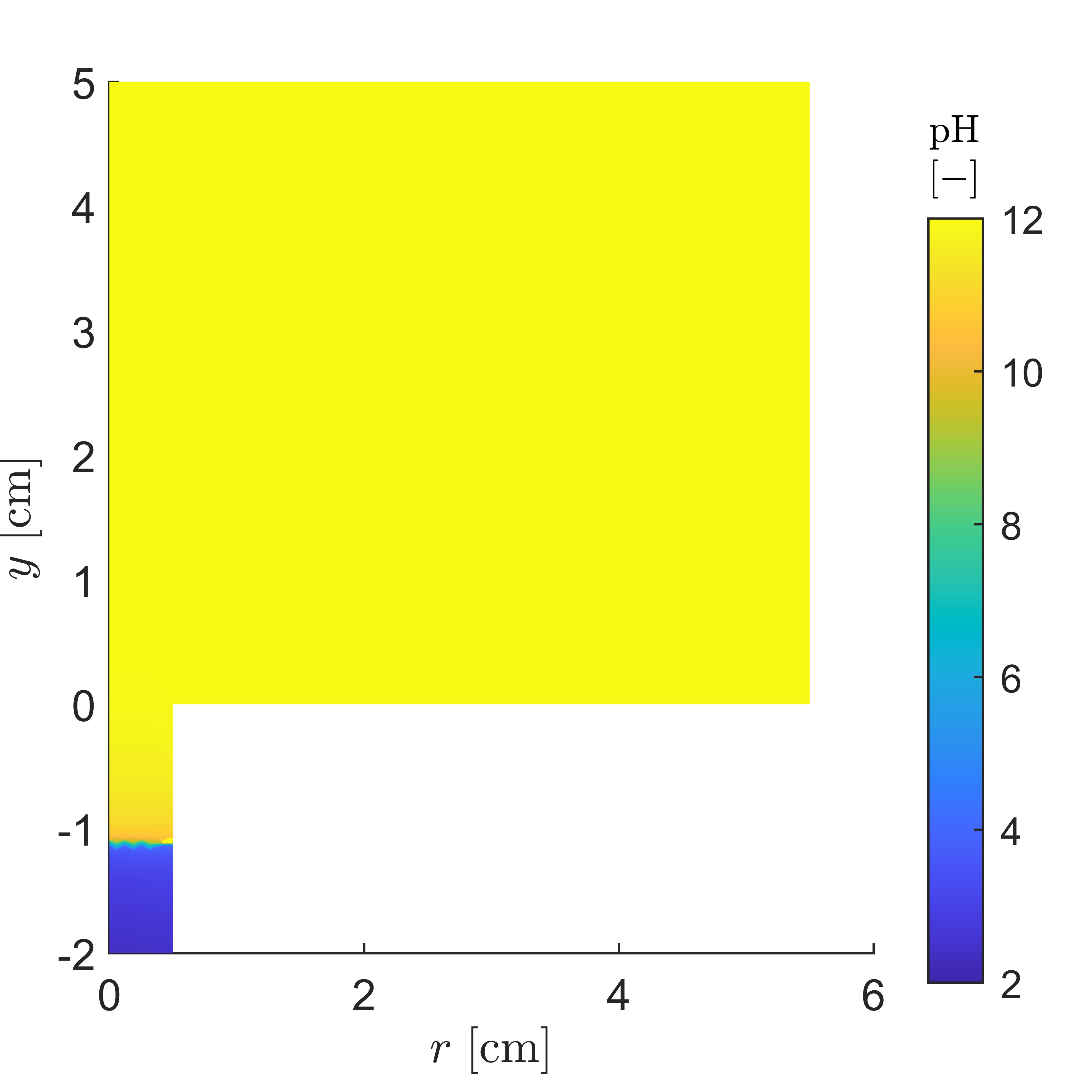}
         \caption{$C_{\mathrm{O}_2}=1\;\mathrm{mol}/\mathrm{m}^3$}
         \label{fig:case1_pH_end_7}
     \end{subfigure}
    \caption{Contours of $\mathrm{pH}$ after $7\;\mathrm{days}$ for oxygen concentrations $C_{\mathrm{O}_2}=0.1\;\mathrm{mol}/\mathrm{m}^3$ (a) and $C_{\mathrm{O}_2}=1\;\mathrm{mol}/\mathrm{m}^3$ (b). Case study A, for a height of $h_{\mathrm{domain}}=5\;\mathrm{cm}$.}
    \label{fig:case1_pH_end}
\end{figure}
\begin{figure}
    \centering
    \begin{subfigure}{8cm}
         \centering
         \includegraphics[width=8cm]{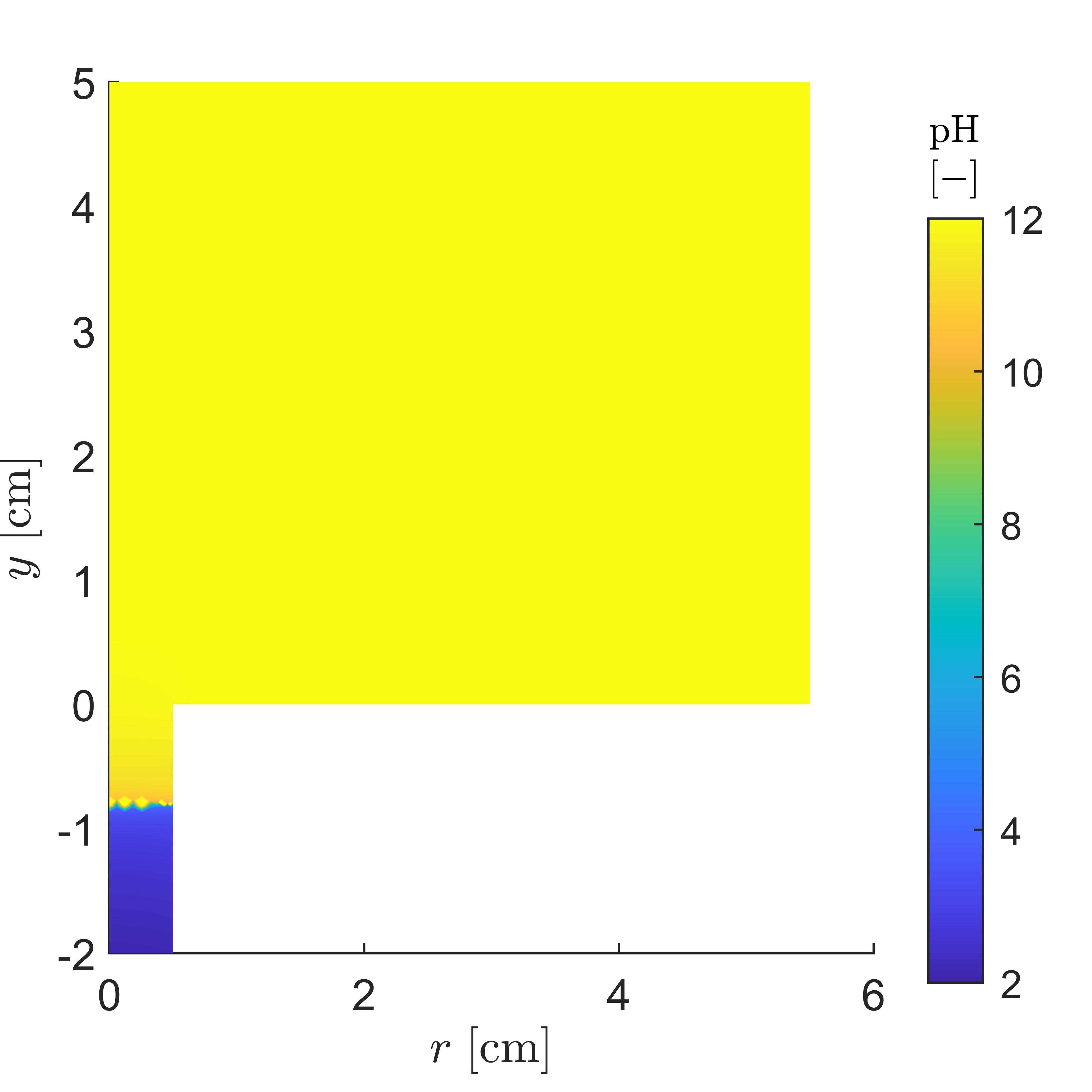}
         \caption{$t=16\;\mathrm{hours}$}
         \label{fig:case1_pH_6a}
    \end{subfigure}
    \begin{subfigure}{8cm}
         \centering
         \includegraphics[width=8cm]{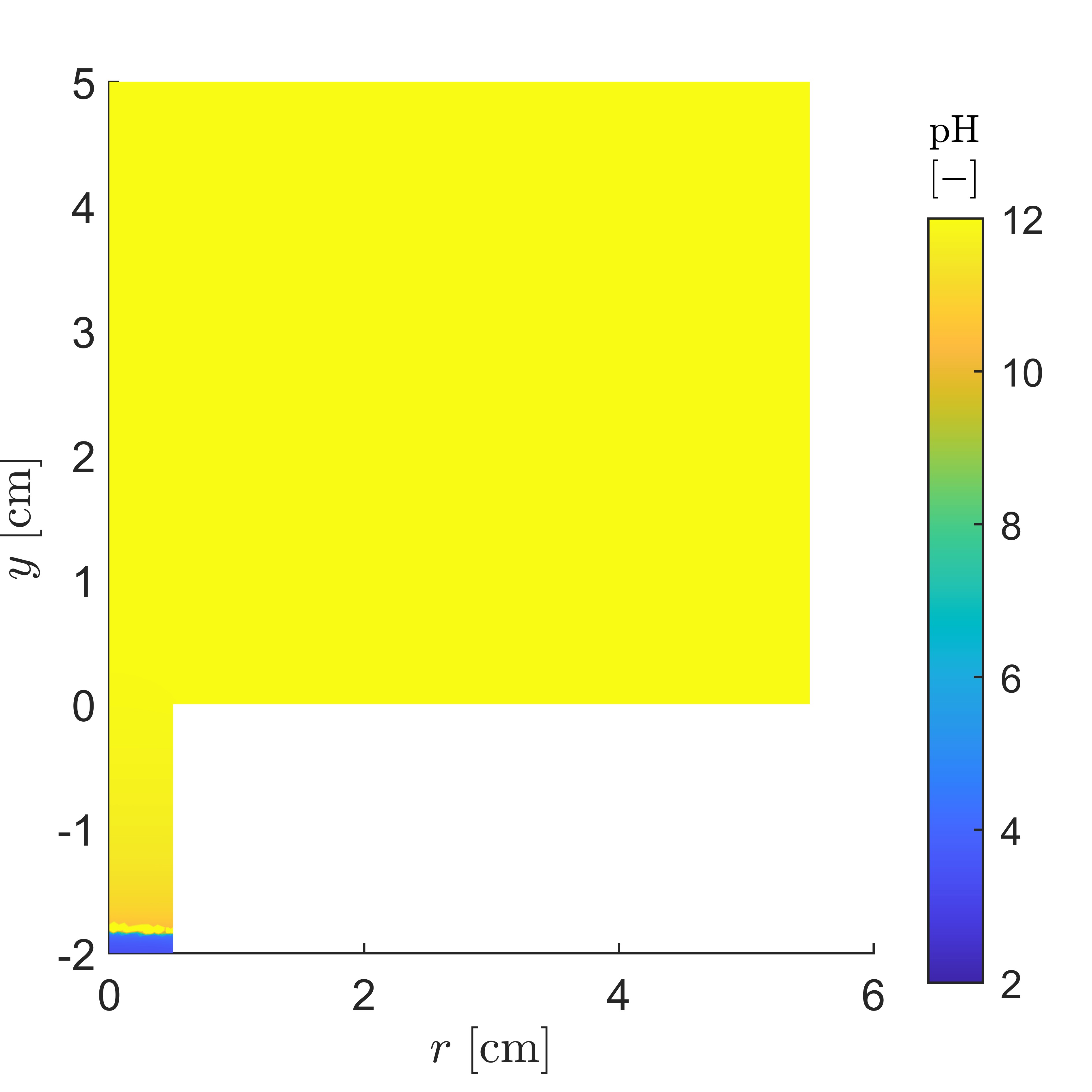}
         \caption{$t=126\;\mathrm{hours}$}
         \label{fig:case1_pH_6b}
     \end{subfigure}
    \caption{Contours of $\mathrm{pH}$ after 16 hours (a), and after 126 hours (b). Case study A, for a height of $h_{\mathrm{domain}}=5\;\mathrm{cm}$ and an oxygen concentration of $C_{\mathrm{O}_2}=0.32\;\mathrm{mol}/\mathrm{m}^3$}
    \label{fig:case1_pH_6}
\end{figure}

\begin{figure}
    \centering
    \begin{subfigure}{8cm}
         \centering
         \includegraphics[width=8cm,clip,trim={0 150 0 0}]{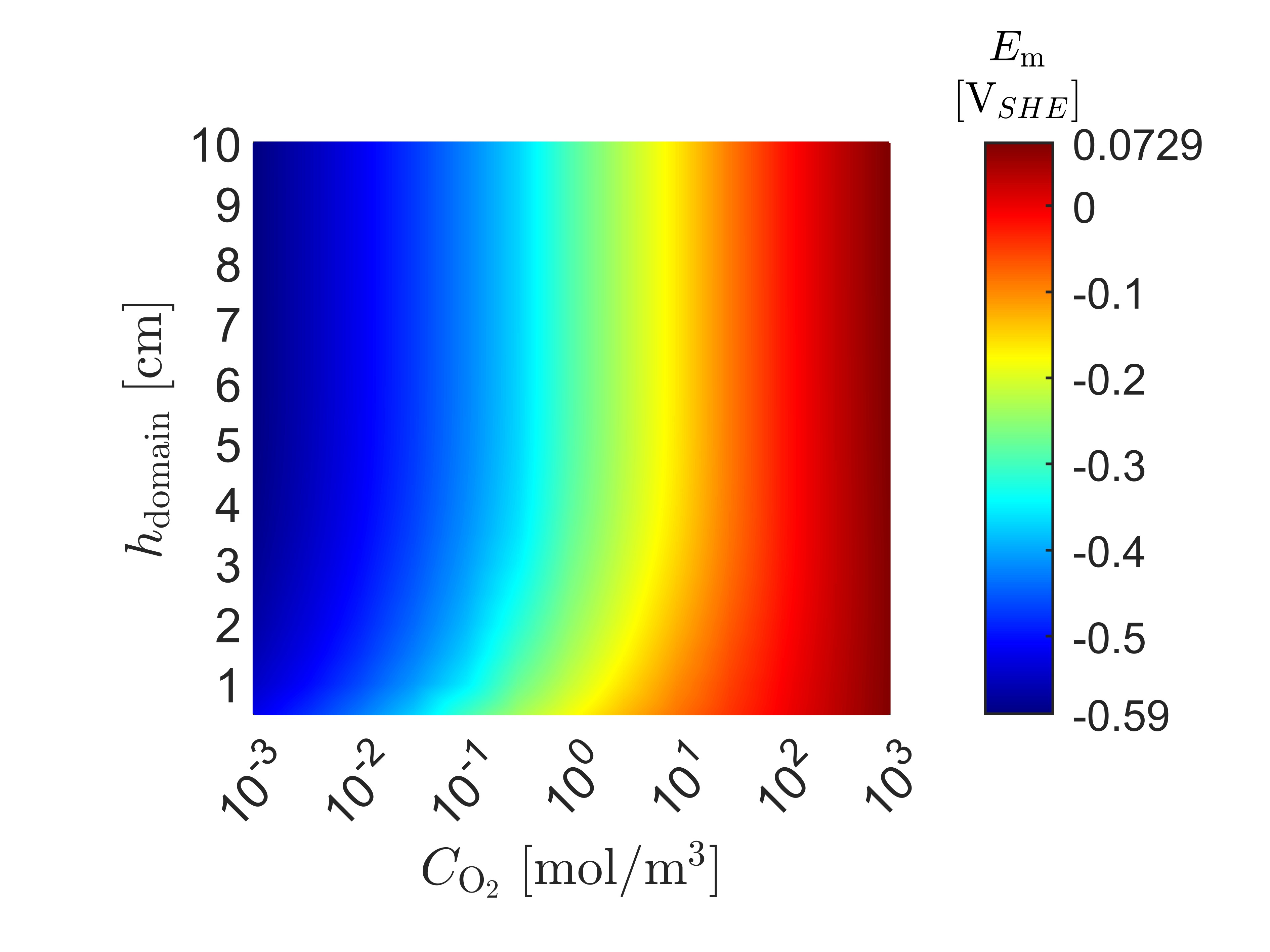}
         \caption{ }
         \label{fig:case1_surfs_Em}
    \end{subfigure}
    \begin{subfigure}{8cm}
         \centering
         \includegraphics[width=8cm,clip,trim={0 150 0 0}]{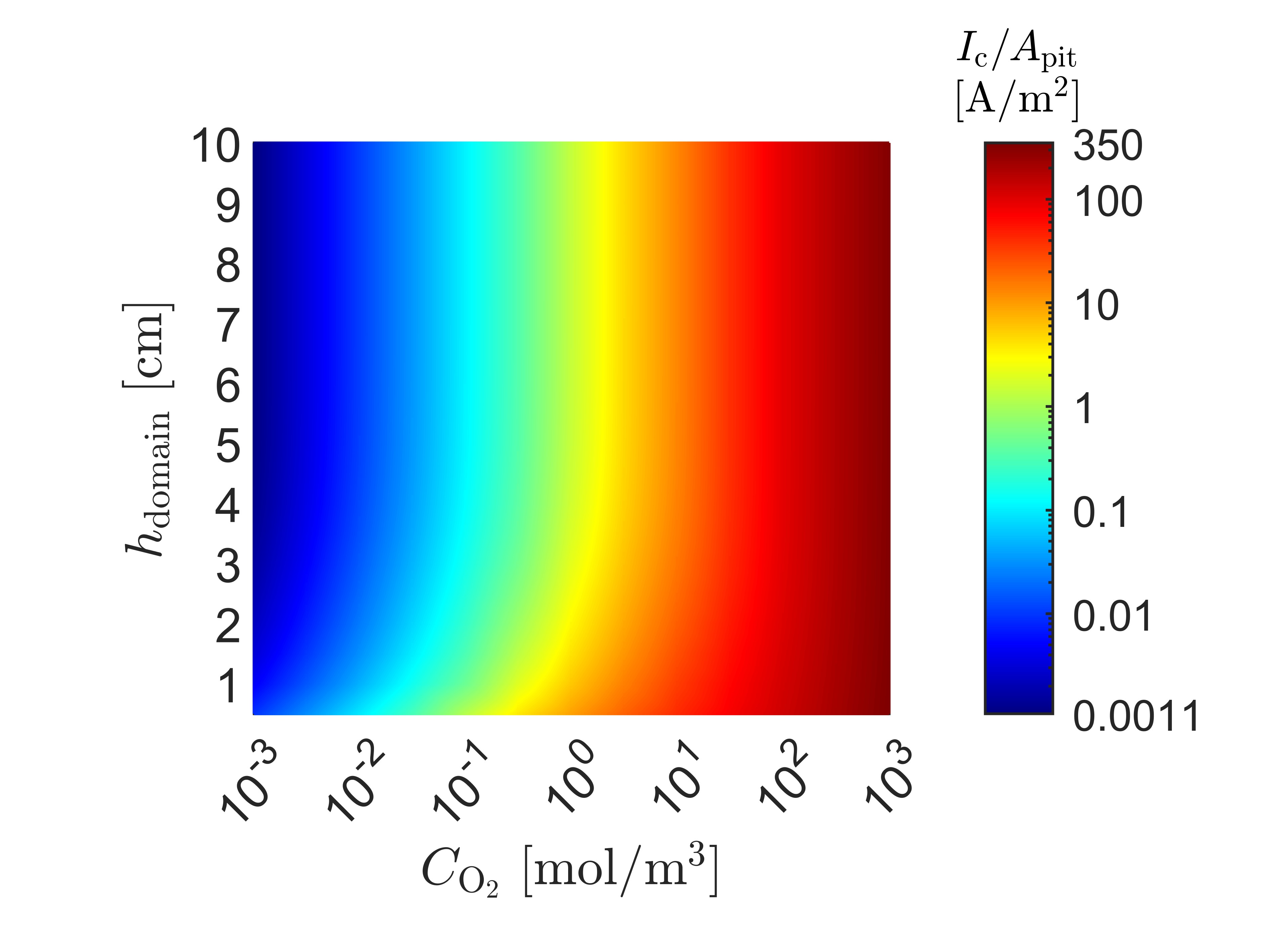}
         \caption{ }
         \label{fig:case1_surfs_Corr}
     \end{subfigure}
     \begin{subfigure}{8cm}
         \centering
         \includegraphics[width=8cm,clip,trim={0 150 0 0}]{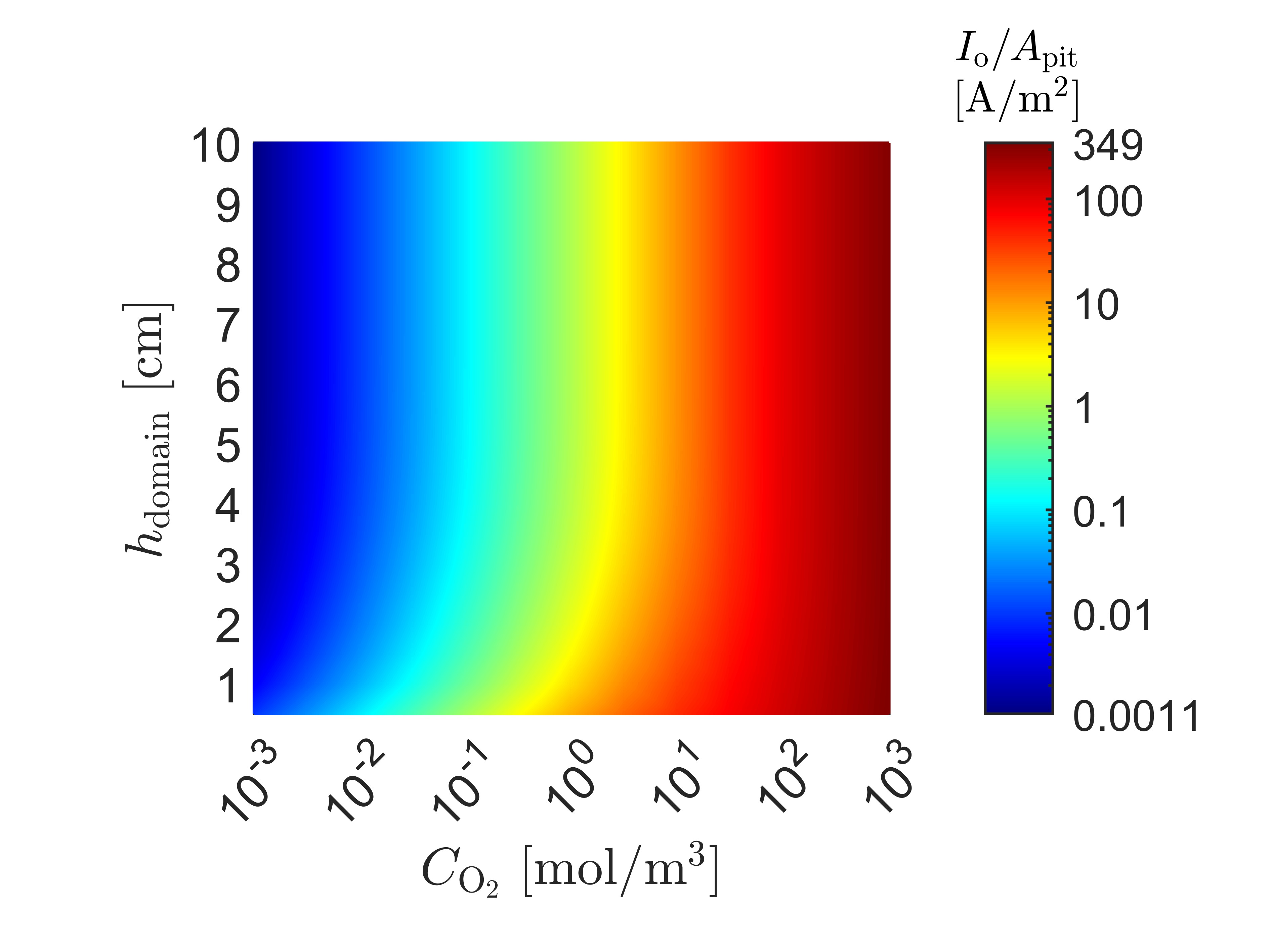}
         \caption{ }
         \label{fig:case1_surfs_O2}
    \end{subfigure}
    \begin{subfigure}{8cm}
         \centering
         \includegraphics[width=8cm,clip,trim={0 150 0 0}]{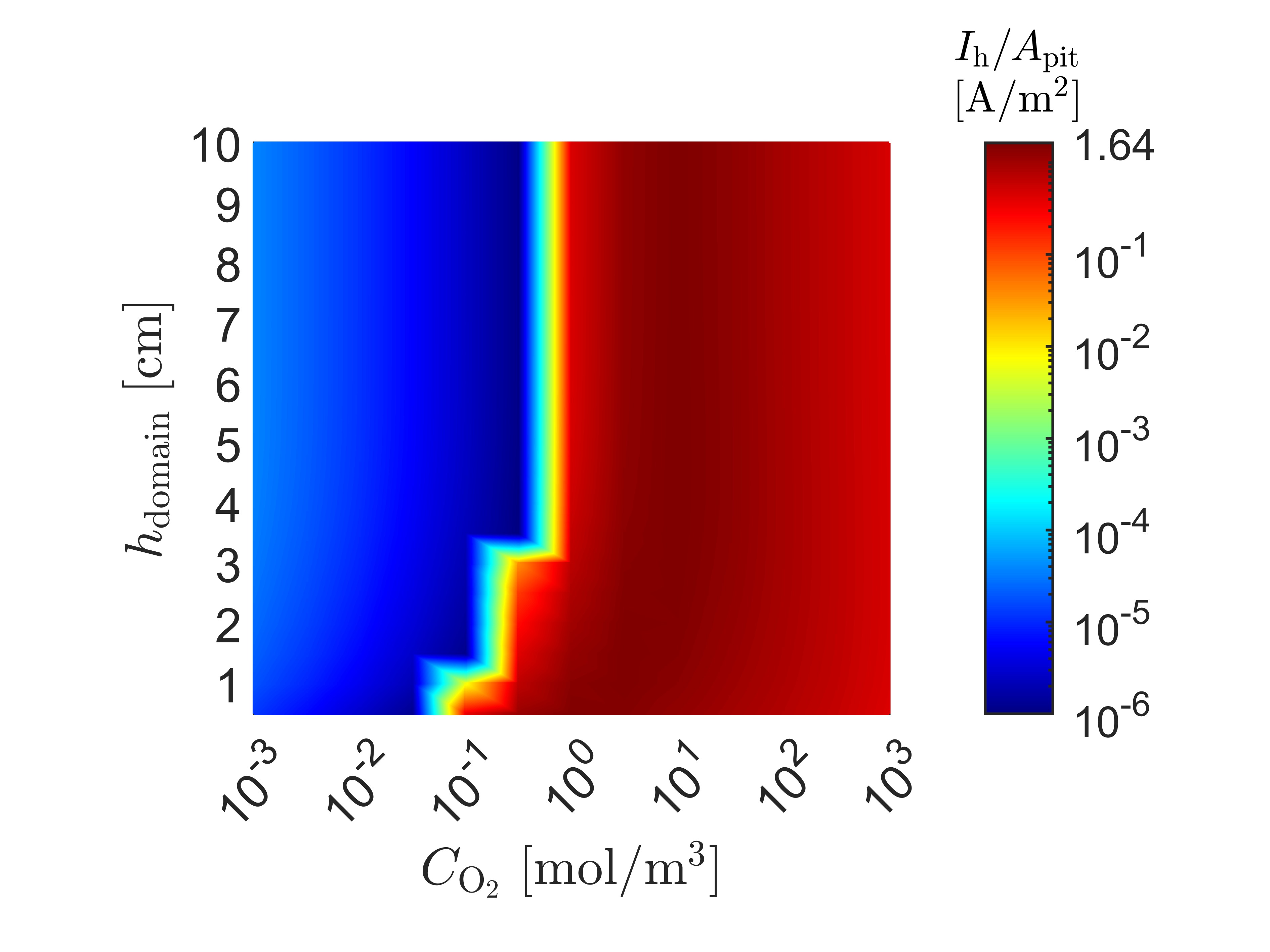}
         \caption{ }
         \label{fig:case1_surfs_H}
     \end{subfigure}
    \caption{Case study A: Mapping the influence of oxygen concentration and domain height on the metal potential (a), the corrosion current (b), the oxygen reaction current (c), and the hydrogen reaction current (d). Results obtained for a time of 7 days.}
    \label{fig:case1_surfs}
\end{figure}
\subsection{Case study A: Role of oxygen concentration and domain height}

The corrosion, hydrogen, and oxygen reaction currents for the first case are shown in \cref{fig:case1_EM_Current} using $h_{\mathrm{domain}}=5\;\mathrm{cm}$ and a wide range of oxygen concentrations. Comparing the hydrogen (Fig. \ref{fig:case1_EM_Current}d) and oxygen (Fig. \ref{fig:case1_EM_Current}d) related currents shows that for all cases, even for the lowest oxygen concentrations, the oxygen current is dominant. This current is balanced out by the corrosion current (Fig. \ref{fig:case1_EM_Current}b), with the electric potential of the metal changing to balance these two contributions (Fig. \ref{fig:case1_EM_Current}a). As shown in Fig. \ref{fig:case1_EM_Current}a, when a low oxygen concentration is present, the metal potential becomes highly negative to reduce the corrosion rate, limiting the amount of electrons produced by the corrosion reaction to a level that can be consumed by the oxygen surface reaction. In contrast, for high oxygen concentrations the metal potential increases due to the increased oxygen reaction current, requiring the corrosion reaction to be accelerated to supply enough electrons to sustain the oxygen reaction. As a result, both the corrosion and oxygen reaction currents are low for low oxygen contents, while increasing the amount of oxygen at the domain boundary increases the metal potential and the corrosion rate inside the pit, scaling as $\log(C_{\mathrm{O}_2}) \propto E_\mathrm{m} \propto  \log(i_{\mathrm{c}})$.

One interesting observation is the clear division of the hydrogen reaction current results into two ``bands" around $i_\mathrm{h}=1\;\mathrm{A}/\mathrm{m}^2$ and $i_\mathrm{h}=10^{-6}\;\mathrm{A}/\mathrm{m}^2$, with no predictions falling in-between these two distinct regimes at equilibrium. Furthermore, the hydrogen reaction current in the simulation using $C_{\mathrm{O}_2}=0.32\;\mathrm{mol}/\mathrm{m}^3$ is initially located in the upper band, and over time decreases and stabilises in the lower band. This localisation of the results over two distinct regions can be explained by considering the pH contours given in \cref{fig:case1_pH_end}. When a low concentration of oxygen is present (Fig. \ref{fig:case1_pH_end}a), the corrosion rate is severely reduced, and as a consequence the amount of iron ions within the pit reacting to produce $\mathrm{H}^+$ is low, limiting the effect of the pit on the pH. This highly basic environment corresponds to the lower band of hydrogen reaction currents, with these currents being significantly lower compared to the corrosion and oxygen reaction currents. Within this band of hydrogen reaction currents, a spread is created due to changes in metal potential that are associated with changes in oxygen reaction rate. The hydrogen reaction rate increases when the metal potential decreases, making the hydrogen evolution reaction current weakly dependent on the oxygen concentration. When the oxygen concentration increases (Fig. \ref{fig:case1_pH_end}b), so does the corrosion rate, and the pit becomes able to sustain an acidic environment. This acidic environment in turn brings the predicted hydrogen currents to the upper band. Within this band, increased corrosion rates result in a thicker layer of the acidic environment, increasing the hydrogen reaction current. In contrast, increases in metal potential still hinder the hydrogen reaction and thus reduce the reaction current $I_h$. These two effects have an opposite effect, as higher metal potentials both directly slow down the hydrogen evolution reaction rate, and indirectly increase this rate by enhancing corrosion and thickness of the acidic layer. As a result, this upper band is thinner, as increased oxygen availability brings both an increase in metal potential (with the associated decrease in $I_h$) and an acidification of the pit region that increases $I_h$, with these two effects balancing out. Finally, the peculiar behaviour observed for the $C_{\mathrm{O}_2}=0.32\;\mathrm{mol}/\mathrm{m}^3$ simulation, whereby the hydrogen reaction current prediction moves from the upper to the lower band, can be rationalised as follows. First, as shown in Fig. \ref{fig:case1_pH_6}a ($t=16$ h), the initial presence of oxygen in the solution results in the development of an acidic region within the pit. However, as shown in Fig. \ref{fig:case1_pH_6}b ($t=126$ h), as this initial oxygen gets consumed by the cathodic surface reaction, the corrosion rate is no longer sufficient to sustain the acidic region and the hydrogen reaction current decreases to align with the basic pit region.\\

We proceed to map the influence of both oxygen concentration and domain height, see \cref{fig:case1_surfs}. Consider first the results for the hydrogen reaction currents (Fig. \ref{fig:case1_surfs}d); as was the case for $h_{\mathrm{domain}}=5\;\mathrm{cm}$ (\cref{fig:case1_EM_Current}b), predictions are split into two distinct regions, with a discontinuity between them. The finite number of simulations conducted inevitably makes this division between regions rather abrupt but its smoothness improves with increasing number of calculations. Regarding the oxygen reaction current (Fig. \ref{fig:case1_surfs}c), the reaction rate is seen to decrease with increasing domain height and increase with increasing oxygen concentration, with the effect being particularly strong for the latter. This scales with $i_\mathrm{o} \propto C_{\mathrm{O}_2}/h_{\mathrm{domain}}$, indicating that for low oxygen concentrations, the key quantity for the oxygen current is the amount of oxygen diffusing from the prescribed boundary towards the cathodic surface. In contrast, for higher concentrations the surface is unable to react with all the available oxygen, resulting in a lower influence of the domain height while still being strongly dependent on the amount of oxygen available. The results obtained for the corrosion current (Fig. \ref{fig:case1_surfs}b) closely resemble those of the oxygen reaction current as they are coupled through the charge conservation condition and the oxygen reaction determines the limiting rate for corrosion. Finally, the metal potential (Fig. \ref{fig:case1_surfs}a) is seen to change from $-0.59\;\mathrm{V}_{\mathrm{SHE}}$ up to $0.07\;\mathrm{V}_{\mathrm{SHE}}$ to balance the anodic and cathodic reactions. This indicates that the lack of oxygen near the metal surface results in a lower metal potential and thus inhibits corrosion, whereas a close-by source of oxygen can increase the corrosion rate by orders of magnitude.

\subsection{Case study B: Role of domain height and pit depth}
\begin{figure}
    \centering
    \begin{subfigure}{8cm}
         \centering
         \includegraphics[width=8cm,clip,trim={0 250 0 0}]{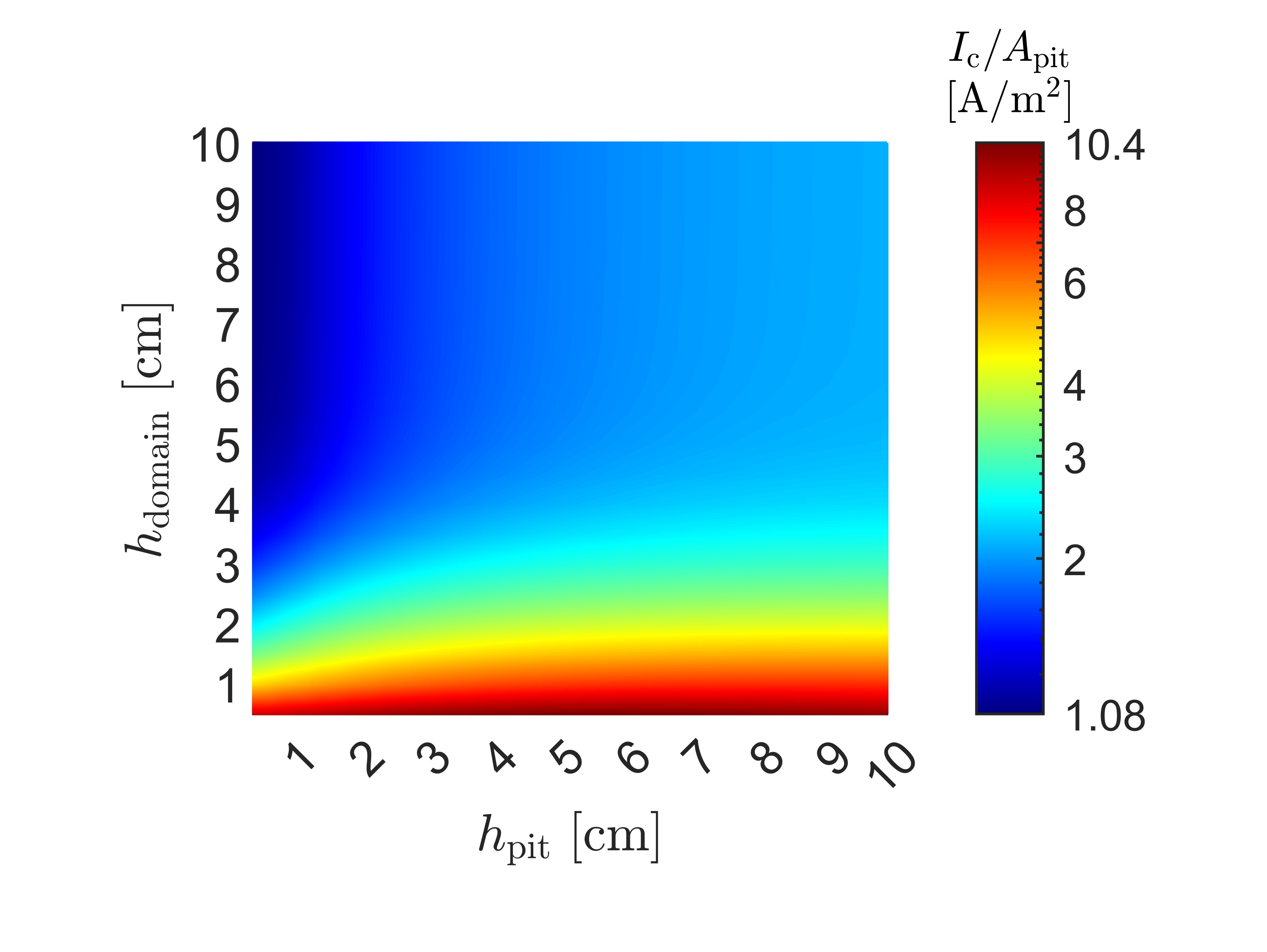}
         \caption{ }
         \label{fig:case2_surfs_Corr}
     \end{subfigure}
     \begin{subfigure}{8cm}
         \centering
         \includegraphics[width=8cm,clip,trim={0 250 0 0}]{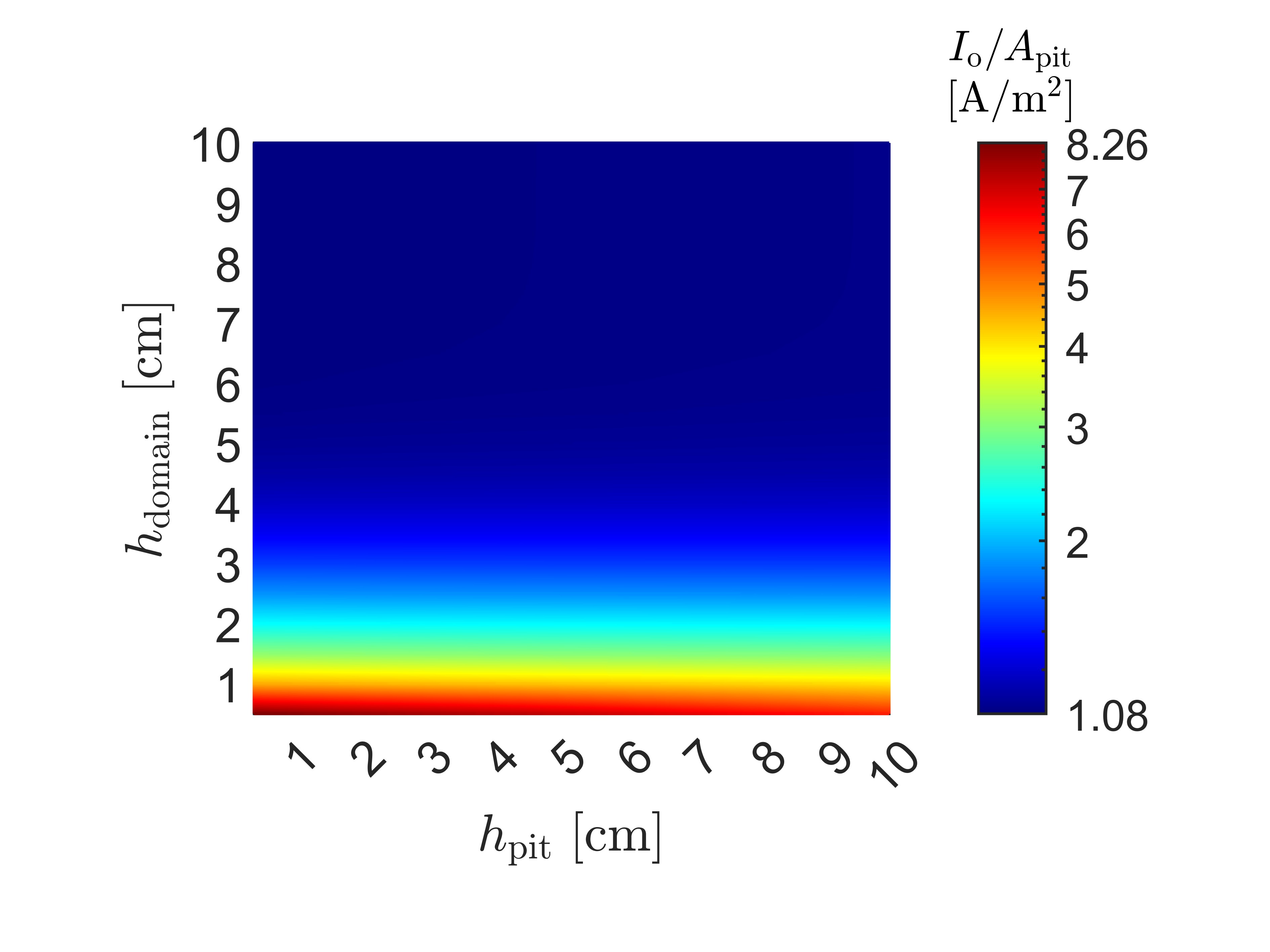}
         \caption{ }
         \label{fig:case2_surfs_O2}
    \end{subfigure}
    \caption{Case study B: Mapping the influence of the domain height $h_{\mathrm{domain}}$ and the pit depth $h_{\mathrm{pit}}$ on the corrosion current (a) and the oxygen reaction current (b). Results obtained for a time of 7 days.}
    \label{fig:case2_surfs}
\end{figure}

In the second case study, case study B, we examine the influence of varying the domain height and pit depth; the results obtained are given in Fig. \ref{fig:case2_surfs}. The results obtained for the hydrogen reaction current are not shown, as the predictions are similar to those of case study A; the hydrogen reaction rate is negligible when only a small amount of corrosion occurs, and nearly independent of pit and domain height when the low pH region can sustain itself. The results for the rate of the oxygen reaction are given in Fig. \ref{fig:case2_surfs}b. As it can be seen, the oxygen reaction is independent of the pit height, and the effect of the domain height is comparable to the diffusion-limited transport behaviour reported in case study A. The hydrogen and oxygen reactions determine the corrosion reaction behaviour (Fig. \ref{fig:case2_surfs}a) and, consequently, the highest corrosion rate is obtained for small domain heights, where the rate is dictated by the oxygen reaction. For large heights, the hydrogen reaction rate is dominant, resulting in a near to constant corrosion rate that exhibits little sensitivity to pit depth and domain height. 

\subsection{Case study C: Role of the pit dimensions}
\begin{figure}
    \centering
    \begin{subfigure}{8cm}
         \centering
         \includegraphics[width=8cm,clip,trim={0 250 0 0}]{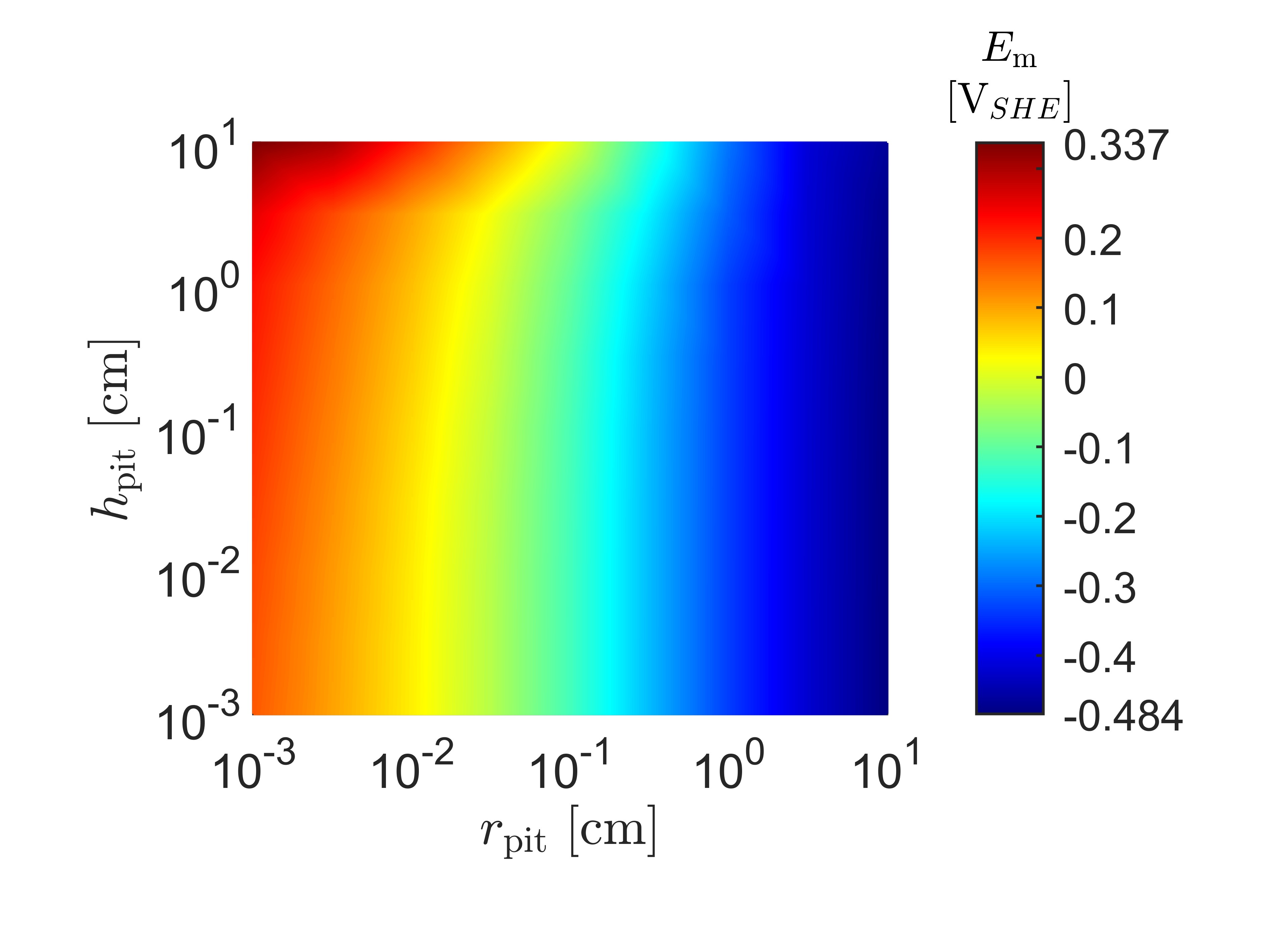}
         \caption{ }
         \label{fig:case3_surfs_Em}
    \end{subfigure}
    \begin{subfigure}{8cm}
         \centering
         \includegraphics[width=8cm,clip,trim={0 250 0 0}]{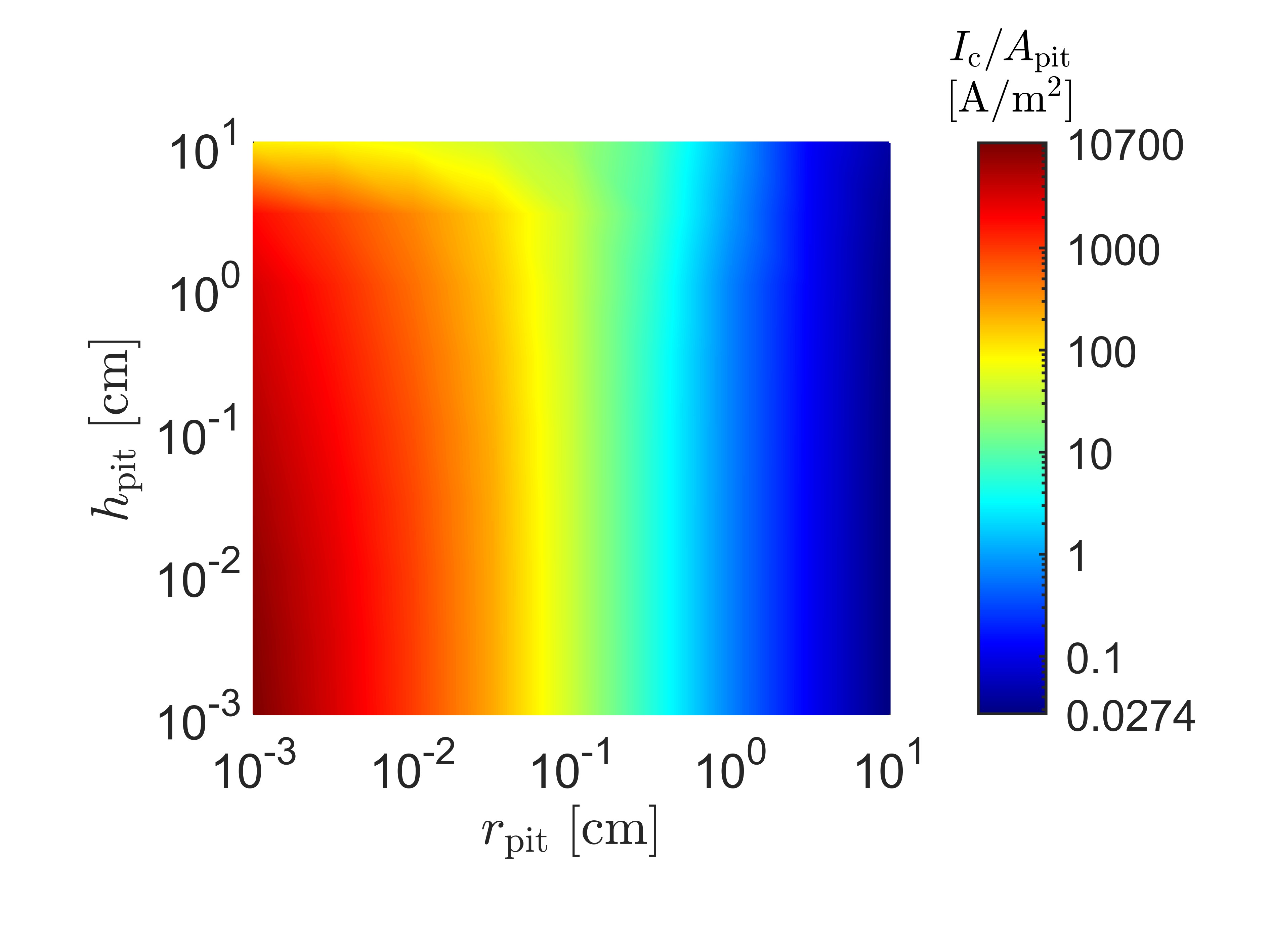}
         \caption{ }
         \label{fig:case3_surfs_O2}
     \end{subfigure}
    \caption{Case study C: Mapping the influence of pit dimension on the metal potential (a) and the corrosion reaction current (b). Results obtained for a time of 7 days.}
    \label{fig:case3_surfs}
\end{figure}
Case study C aims at investigating the role of the pit radius and pit depth, keeping the difference between pit and domain radii constant. The results obtained are shown in \cref{fig:case3_surfs}. As shown in Fig. \ref{fig:case3_surfs}b, increasing the pit radius, and thus the area on which anodic reactions occur, results in lower corrosion currents. The supply of electrons caused by the oxygen reaction on the exterior surface is only dependent on the area of the outer surface. Since this supply limits the corrosion rate, increasing the pit area without significantly changing the outer area does not increase the total corrosion current, resulting in a decrease of the $I_c/A_{\text{pit}}$ value due to the changes in pit area. The results also show that when the pit is small, the local corrosion current is greatly increased. This is a result of the cathodic reactions only needing to balance the anodic reactions of a smaller area, and thus allowing for higher corrosion rates on this small area. To enable these higher corrosion rates required for the charge-conservation condition, the metal potential is increased, accelerating the corrosion reactions while decreasing the cathodic reaction rates. Another limiting factor for small pits is the diffusion towards the exterior of the pit. As the corrosion reaction accelerates, increased electrolyte potentials arise to enforce the electroneutrality condition by forcing the positively charged metal ions to move to the exterior of the pit and attract negative ions towards the pit. Additionally, the increased concentration of iron ions within the corrosion pit allows the corrosion reaction to accelerate the backward corrosion ($\mathrm{Fe}^{2+}$ being reconverted to $\mathrm{Fe}$). These two effects limit corrosion for the smallest and deepest pits simulated, reducing the corrosion rate by several orders of magnitude compared to shallower pits. 

\subsection{Case study D: Role of the pit and domain radii}
\begin{figure}
    \centering
    \begin{subfigure}{8cm}
         \centering
         \includegraphics[width=8cm,clip,trim={0 300 0 0}]{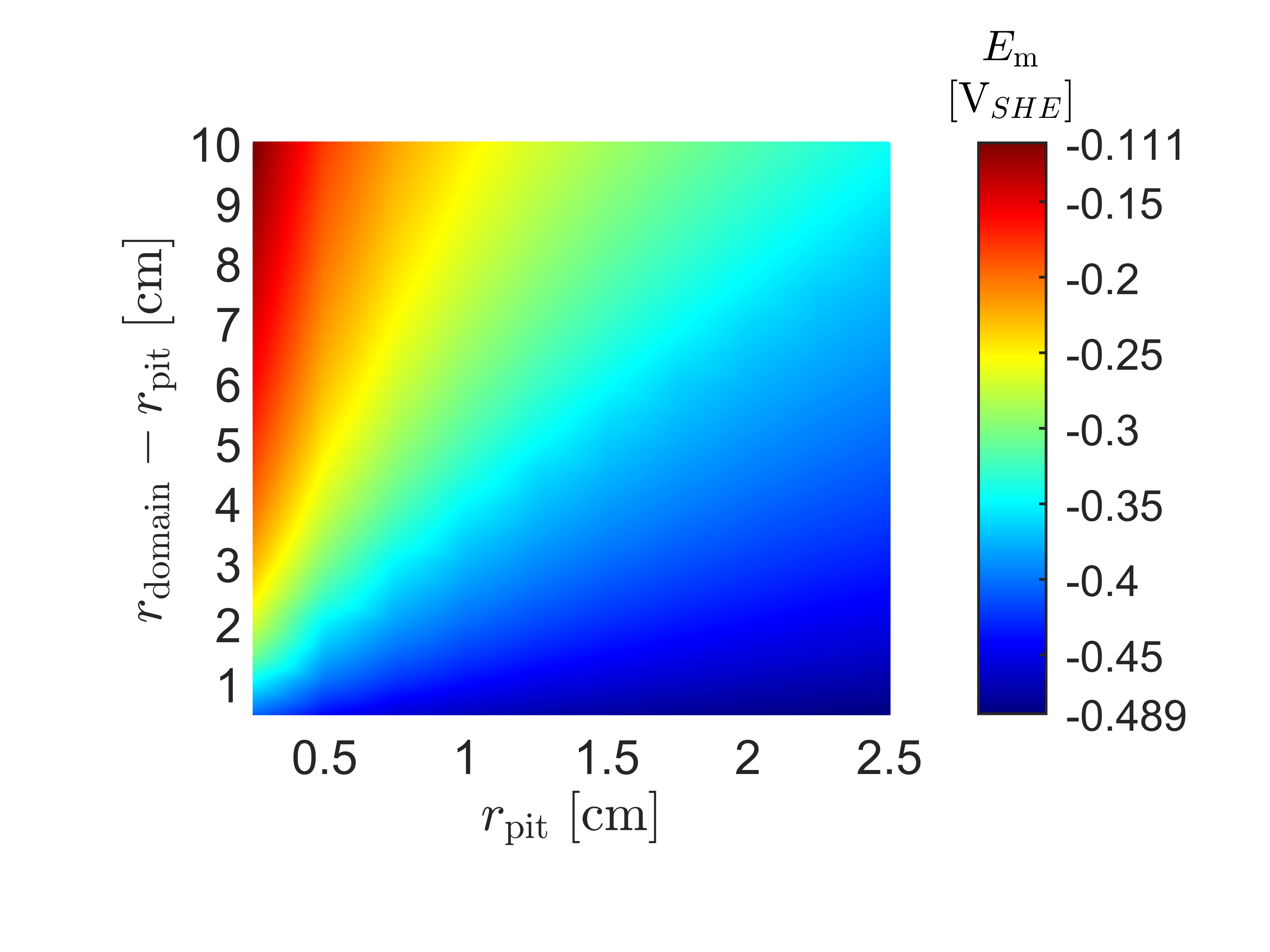}
         \caption{ }
         \label{fig:case4_surfs_Em}
    \end{subfigure}
    \begin{subfigure}{8cm}
         \centering
         \includegraphics[width=8cm,clip,trim={0 300 0 0}]{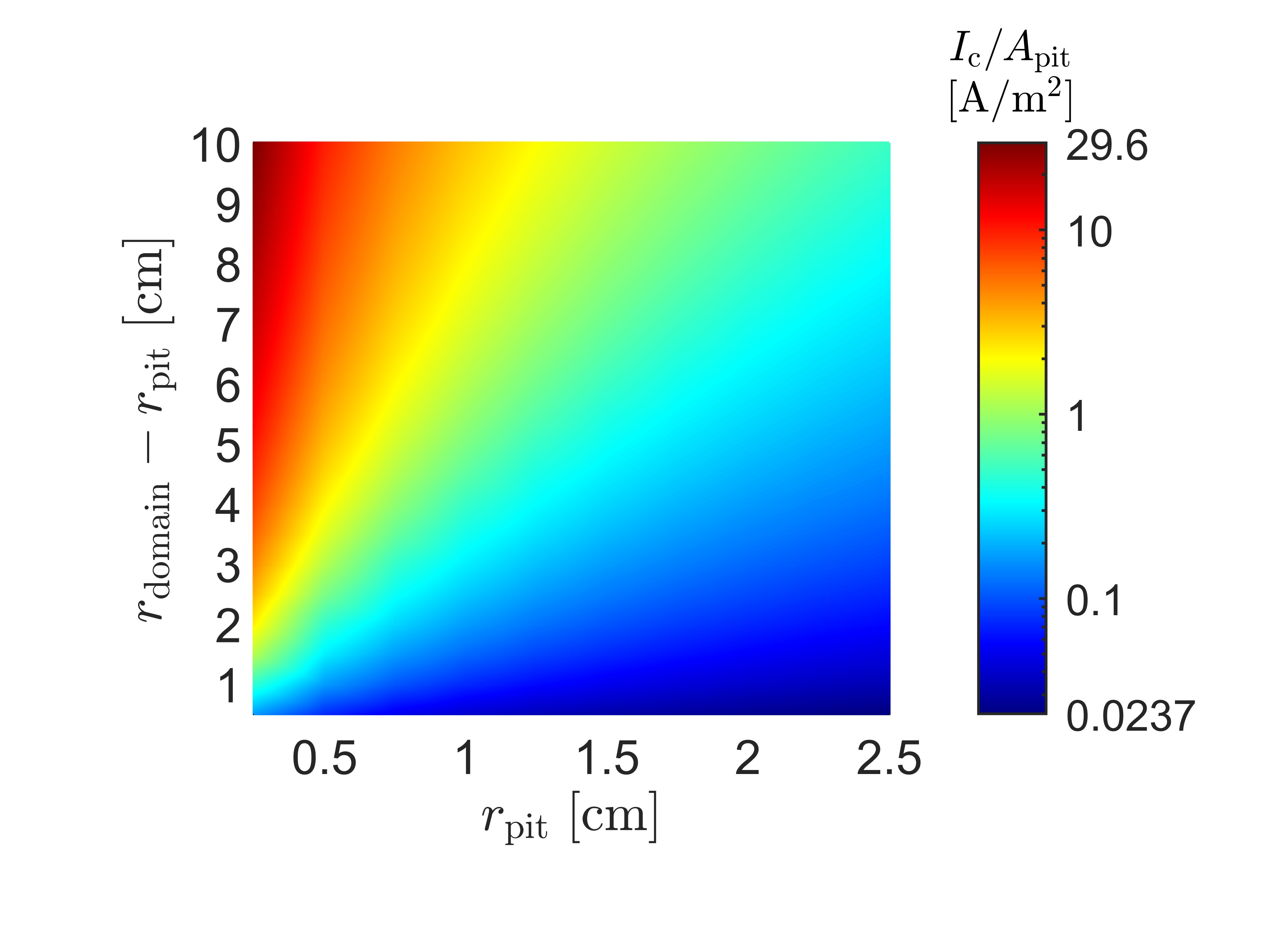}
         \caption{ }
         \label{fig:case4_surfs_Corr}
     \end{subfigure}
     \begin{subfigure}{8cm}
         \centering
         \includegraphics[width=8cm,clip,trim={0 300 0 0}]{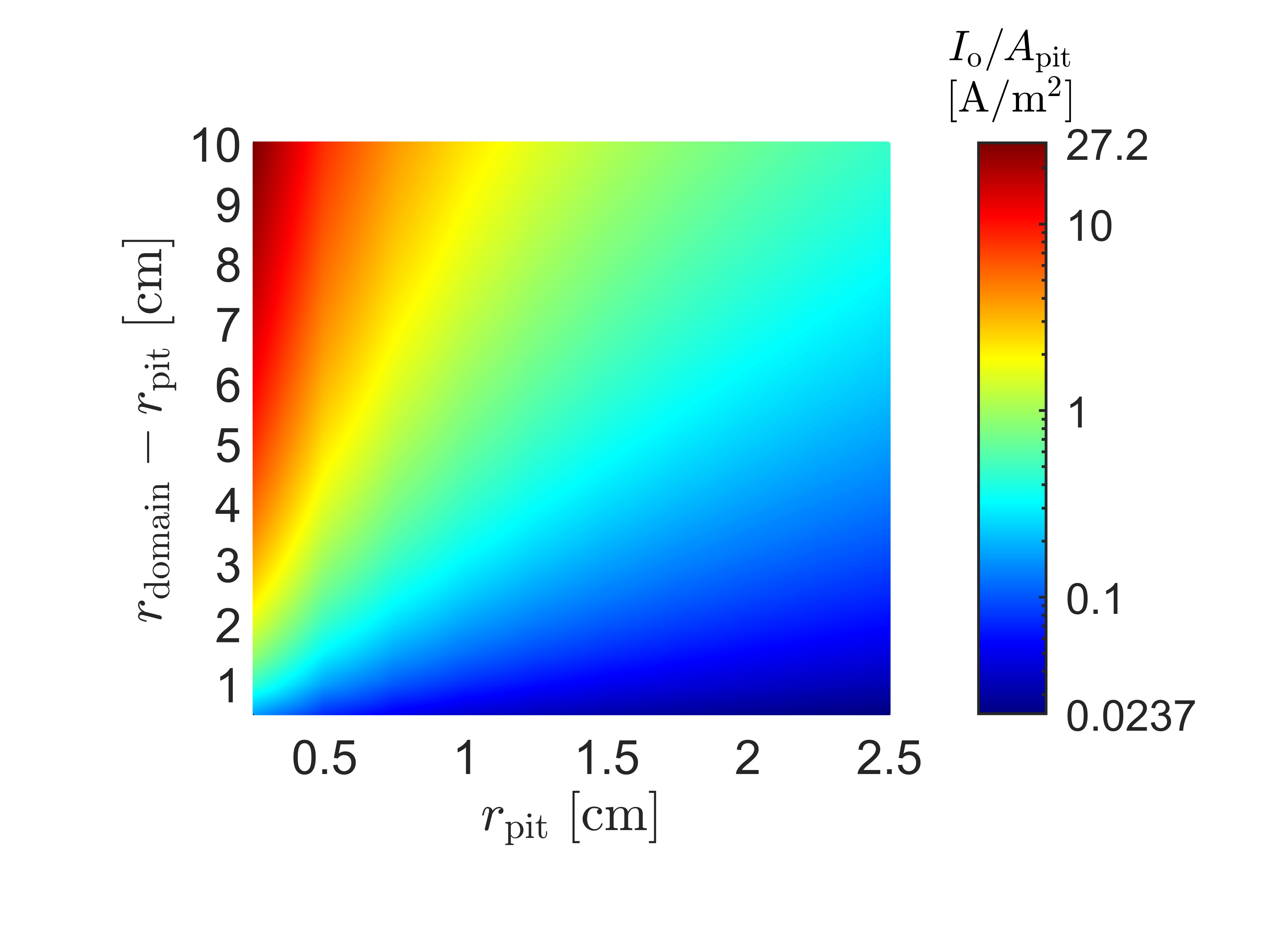}
         \caption{ }
         \label{fig:case4_surfs_O2}
    \end{subfigure}
    \begin{subfigure}{8cm}
         \centering
         \includegraphics[width=8cm,clip,trim={0 300 0 0}]{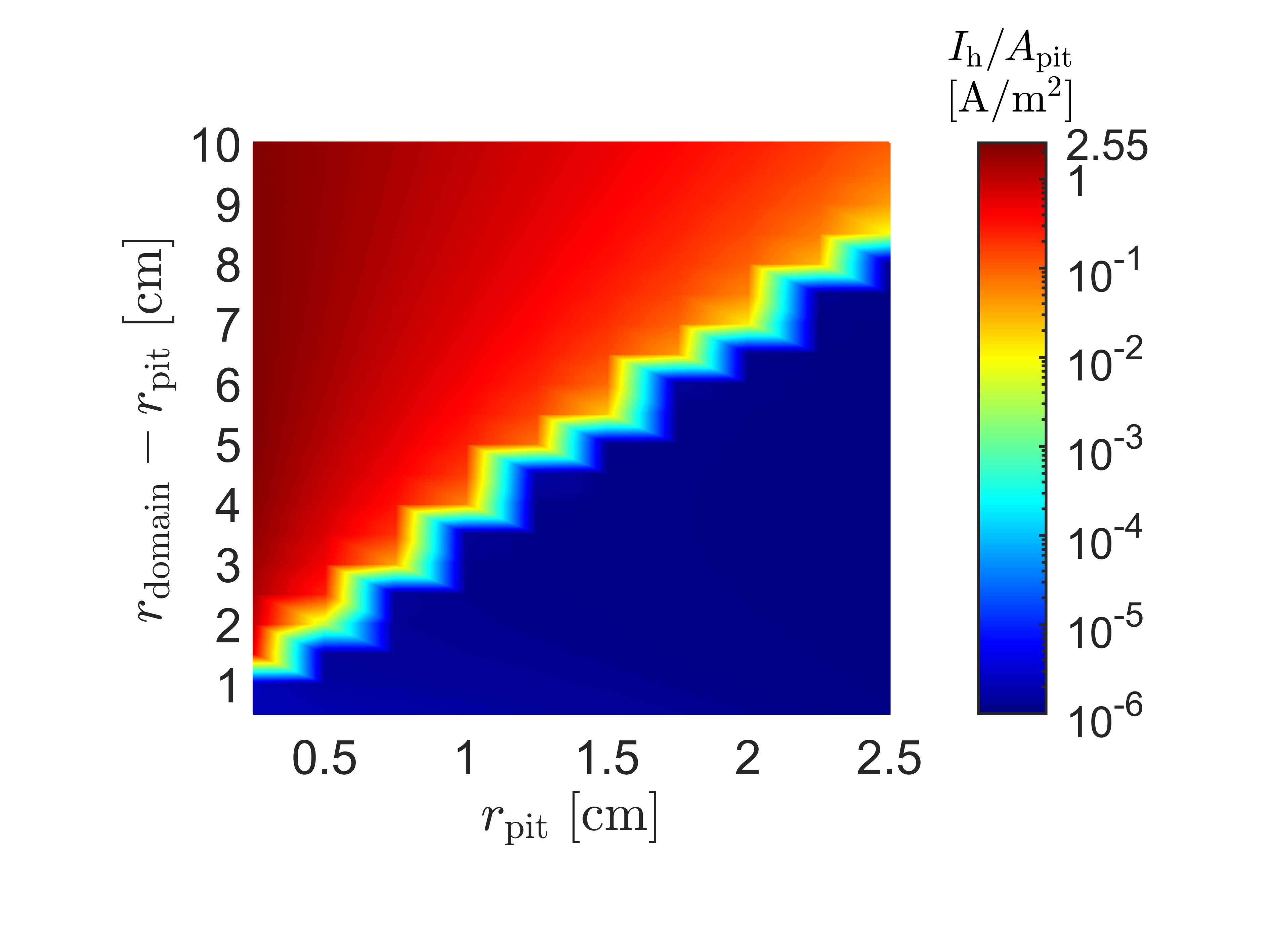}
         \caption{ }
         \label{fig:case4_surfs_H}
     \end{subfigure}
    \caption{Case study D: Mapping the influence of pit and external areas on the metal potential (a), corrosion current (b), oxygen reaction current (c) and hydrogen reaction current (d). Results obtained at a time of 7 days.}
    \label{fig:case4_surfs}
\end{figure}
In case study D, we vary the pit and domain radii to investigate the interactions between external surface area and pit area. The finite element results obtained are given in \cref{fig:case4_surfs}. Consider first the hydrogen reaction current results, Fig. \ref{fig:case4_surfs}d. Similar to case study A, two distinct regions are observed, corresponding to self-sustained acidic and basic pit environments. For wide pits and a relatively small exterior surface area, the corrosion rate is not sufficient to sustain the acidic environment within the pit, as shown in Fig. \ref{fig:case4_surfs}b. The corrosion current increases as the exterior area increases (higher $I_o$, see Fig. \ref{fig:case4_surfs}c) or the pit radius decreases (lower anodic area over which the reaction current can be distributed). This increase in $I_c$ is mostly driven by the oxygen reaction rate, which is an order of magnitude higher compared to the hydrogen reaction current. One thing to note is that the corrosion current is nearly constant for a constant ratio between pit and exterior area radii, indicating that this quantity, and not the radius itself, determines the corrosion rate. Increasing the outer area increases the oxygen reaction rate nearly linearly, and thus the corrosion rate approximately scales with $i_\mathrm{c}\propto r_{\mathrm{domain}}^2/r_{\mathrm{pit}}^2$, only having a small deviation due to the influence of the hydrogen reaction occurring at the pit walls.

\subsection{Case study E: Role of the hydrogen and oxygen reactions}
\begin{figure}
    \centering
    \begin{subfigure}{8cm}
         \centering
         \includegraphics[width=8cm,clip,trim={0 300 0 0}]{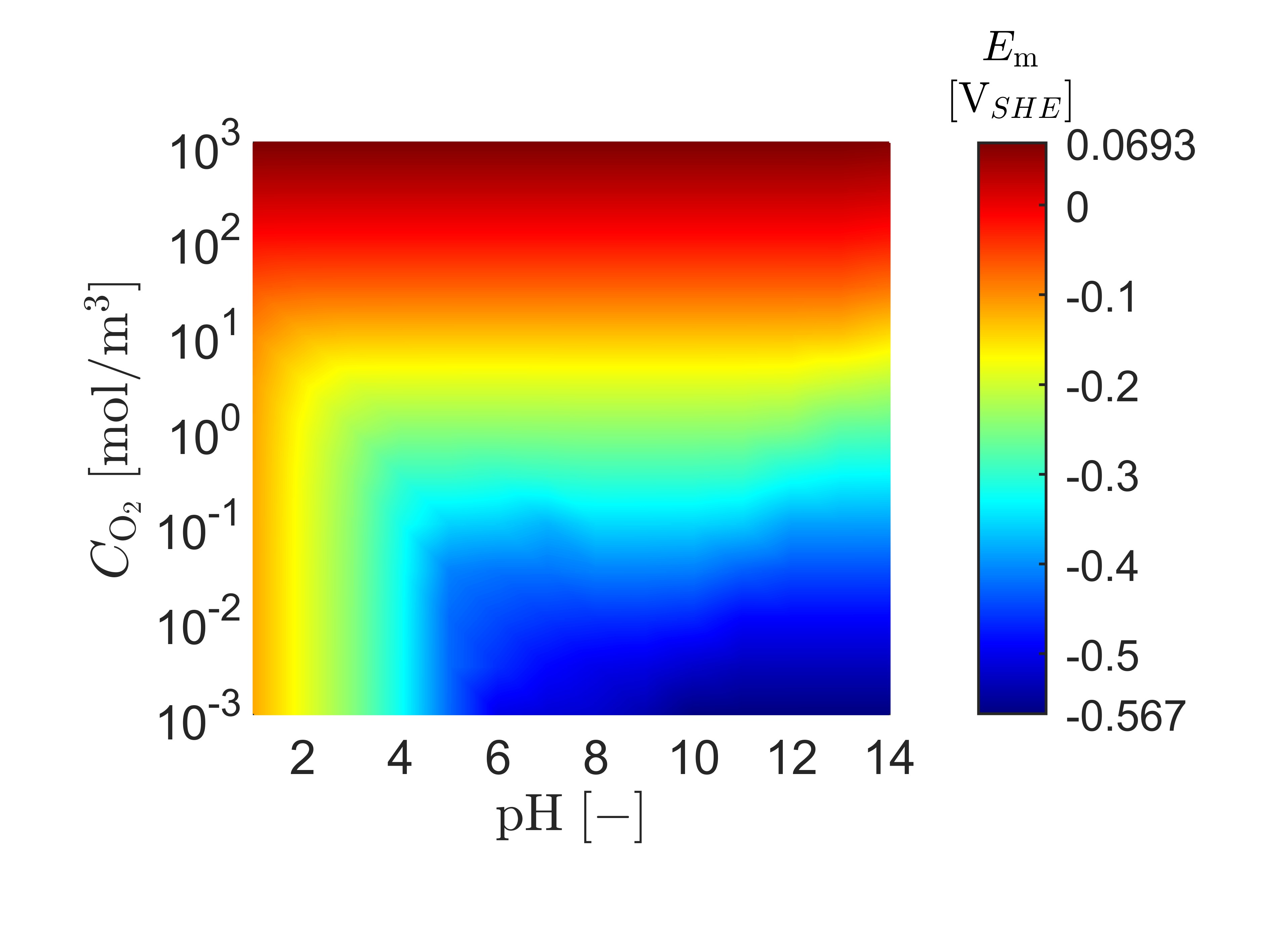}
         \caption{ }
         \label{fig:case5_surfs_Em}
    \end{subfigure}
    \begin{subfigure}{8cm}
         \centering
         \includegraphics[width=8cm,clip,trim={0 300 0 0}]{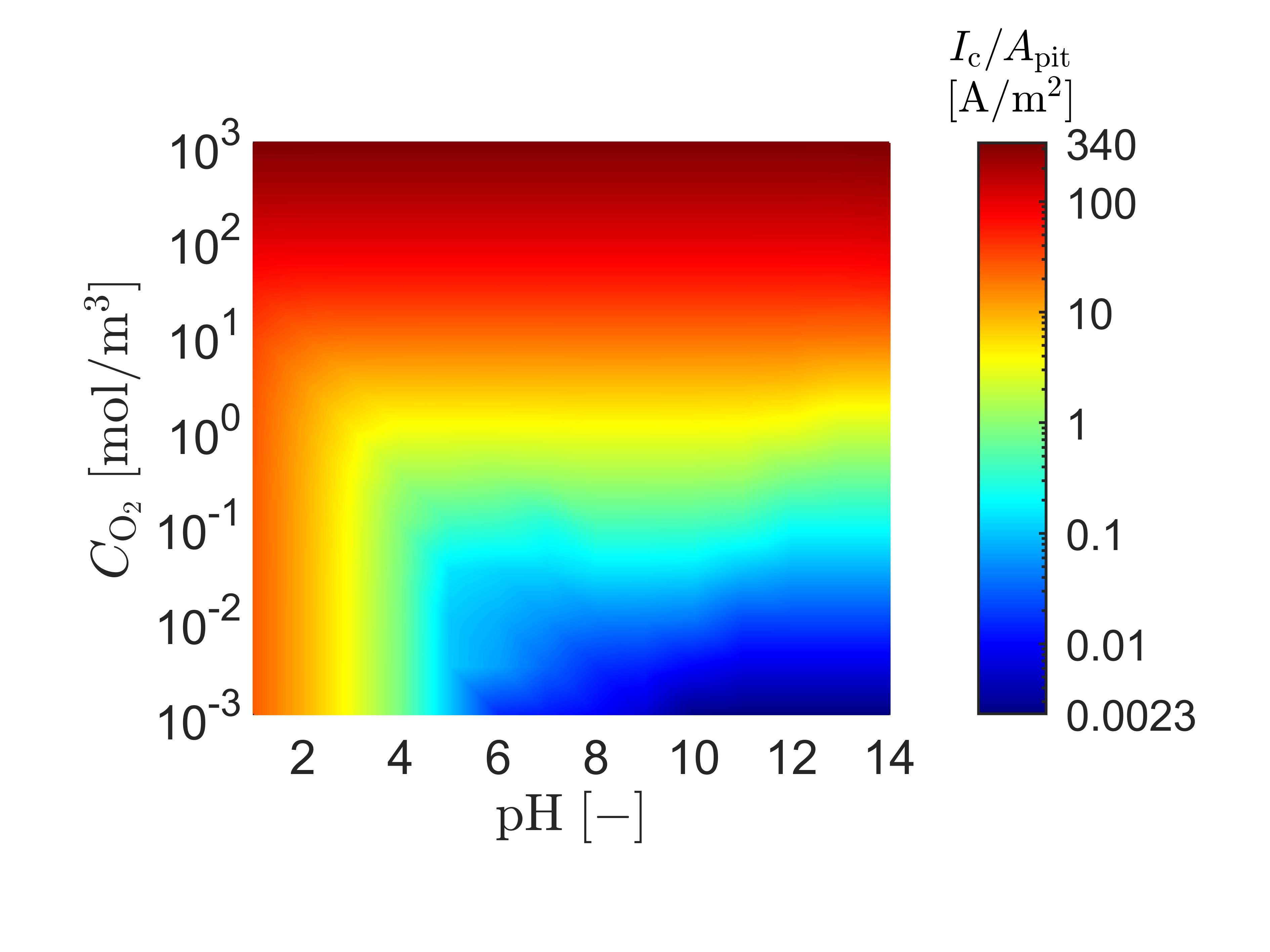}
         \caption{ }
         \label{fig:case5_surfs_Icorr}
     \end{subfigure}
    \begin{subfigure}{8cm}
         \centering
         \includegraphics[width=8cm,clip,trim={0 300 0 0}]{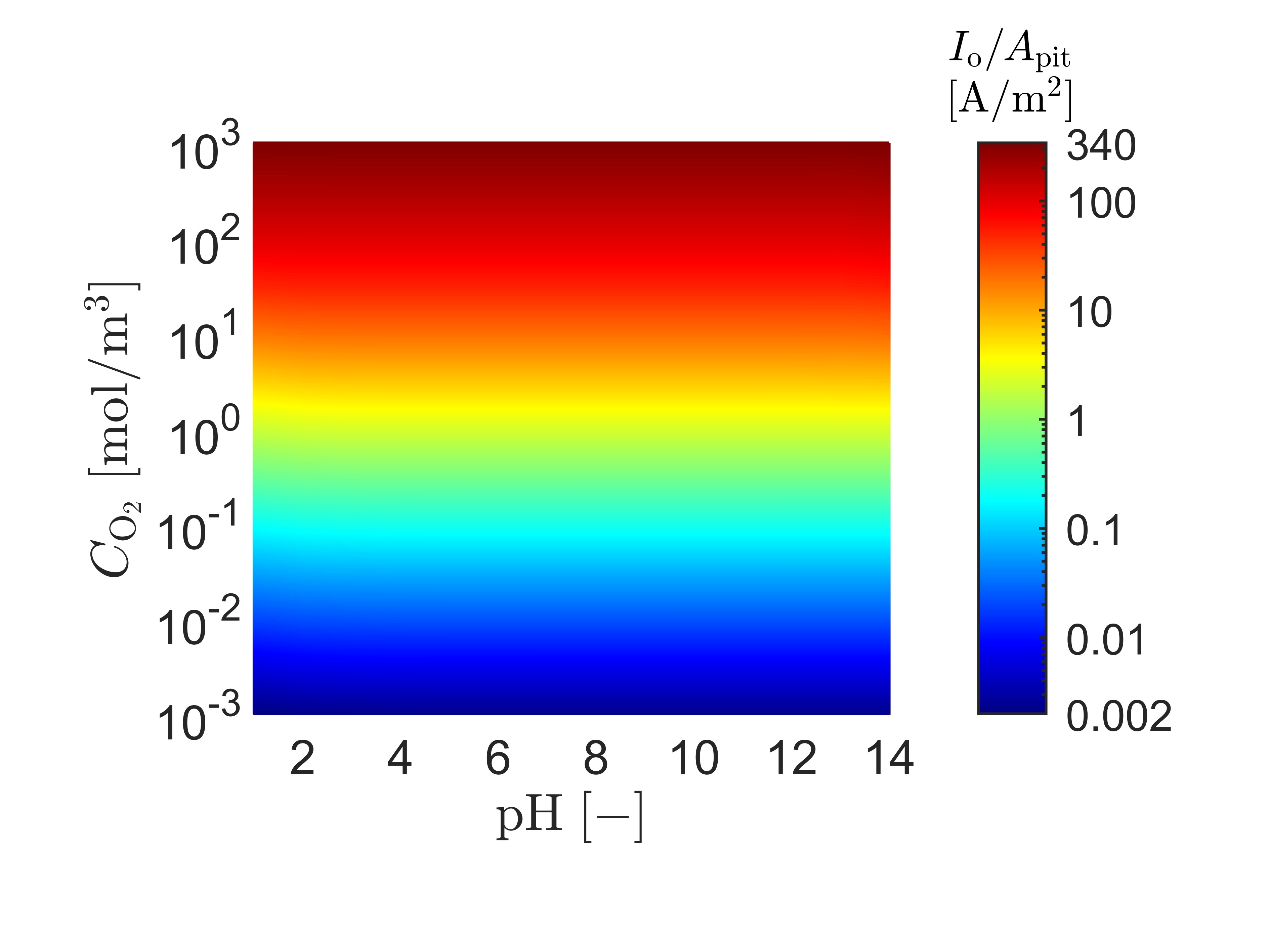}
         \caption{ }
         \label{fig:case5_surfs_O2}
    \end{subfigure}
    \begin{subfigure}{8cm}
         \centering
         \includegraphics[width=8cm,clip,trim={0 300 0 0}]{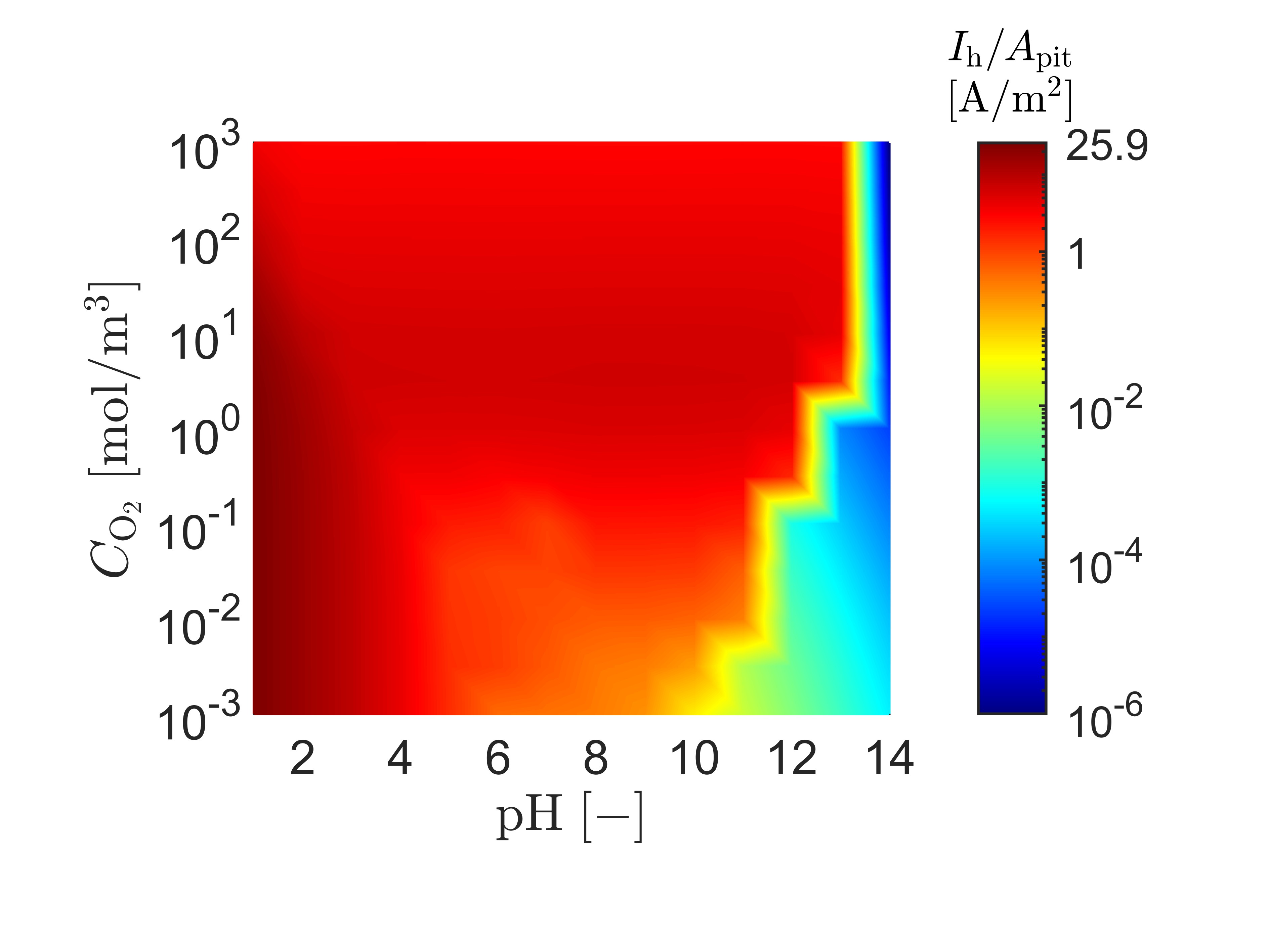}
         \caption{ }
         \label{fig:case5_surfs_H}
     \end{subfigure}
    \caption{Case study E: Mapping the influence of the pH and the oxygen concentration on the metal potential (a), the corrosion current (b), the oxygen reaction current (c) and the hydrogen reaction current (d). Results obtained for a time of 7 days.}
    \label{fig:case5_surfs}
\end{figure}
In case study E, the influence of the pH and oxygen concentration is investigated. To this end, the prescribed quantities at the top of the domain are taken to be the same as those adopted as initial conditions. The results obtained are shown in \cref{fig:case5_surfs}. As observed before, when a high concentration of oxygen is imposed, the oxygen reaction occurs fast enough to sustain a high corrosion rate (see Fig. \ref{fig:case5_surfs}b). Also, in agreement with expectations, the cathodic oxygen current varies with the oxygen concentration, showing little sensitivity to the pH (see Fig. \ref{fig:case5_surfs}c). The hydrogen reaction current results exhibit a more interesting pattern, Fig. \ref{fig:case5_surfs}d, and play a notable role on the resulting corrosion currents. For high $\mathrm{H}^+$ concentrations (low pH), the relatively high values of $I_h$ allow for a significant corrosion current, independently of the oxygen content. However, unlike the oxygen reaction, the hydrogen reaction does not just depend on the pH. Increasing the oxygen concentration allows the oxygen reaction to produce more $\mathrm{OH}^-$ ions, thereby slightly decreasing the pH and the hydrogen reaction current. Furthermore, a sharp change in hydrogen reaction current is observed for higher pH values, with this change being strongest when a large amount of oxygen is present to suddenly trigger the transition to the acidic pit regime. When little oxygen is present, this transition is more gradual, being partly driven by the pH at the exterior surface but also by the pH of the pit being lowered by the initial and boundary conditions imposed. A consequence of the different scaling of the oxygen and hydrogen currents is a large range of pH and oxygen concentrations where corrosion rates are very low. However, as the electrolyte becomes more acidic or more oxygen is present, the corrosion current increases; up to $26\;\mathrm{A}/\mathrm{m}^2$ when the environment is acidic and lacks oxygen (bottom-left region of Fig. \ref{fig:case5_surfs}b), and further up to $340\;\mathrm{A}/\mathrm{m}^2$ when more oxygen is added. 

\subsection{Case study F: Role of the electrolyte conductivity}
\begin{figure}
    \centering
    \includegraphics[width=9cm]{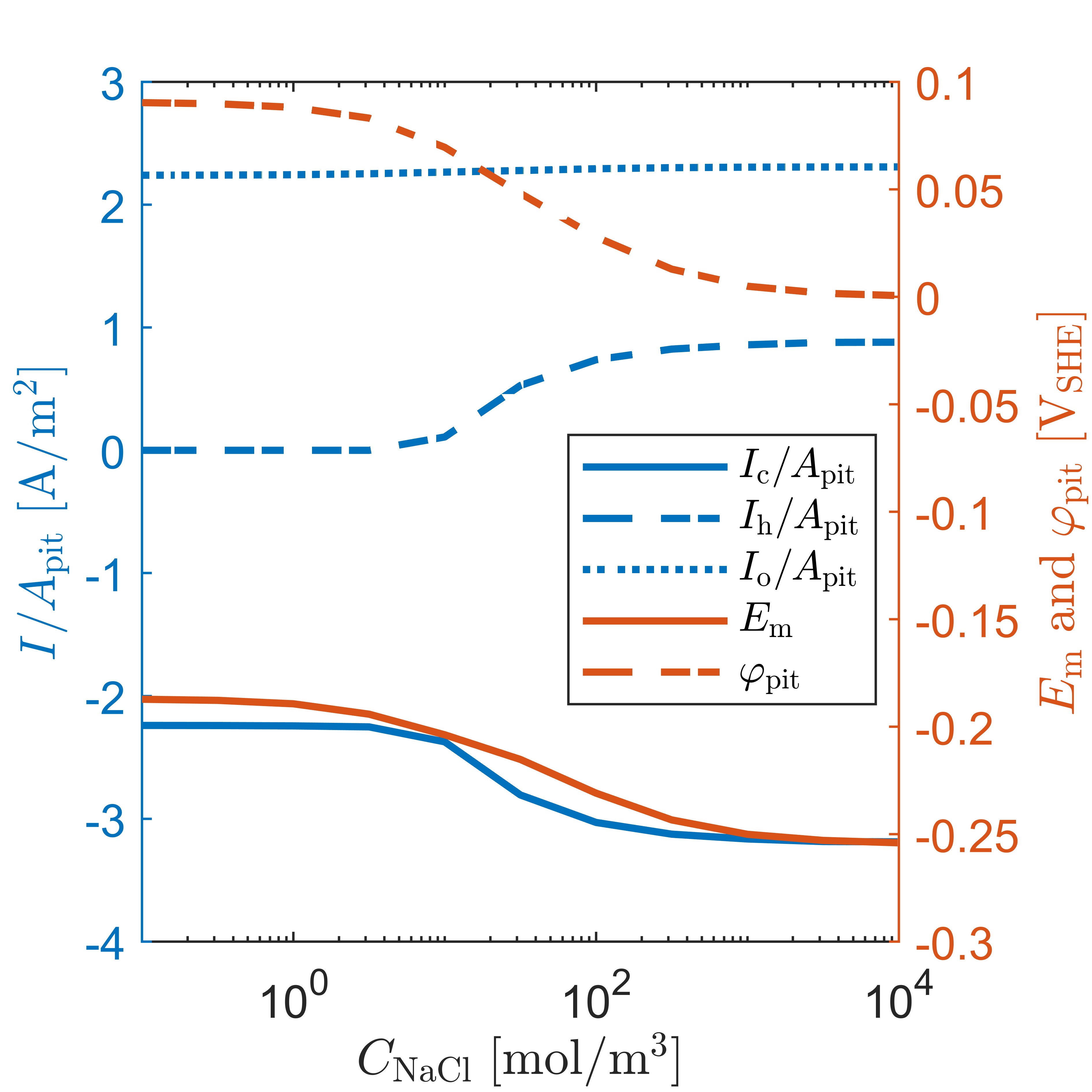}
    \caption{Case study F: influence of the electrolyte conductivity. Predictions of corrosion, hydrogen, and oxygen currents (left axis, blue) and metal and electrolyte potentials (right axis, red). Results obtained for varying $\mathrm{Na}^+$ concentrations and for a time of 7 days.}
    \label{fig:case6_EI}
\end{figure}
The final case study analyses the influence of the electrolyte conductivity. To this end, the electrolyte conductivity is raised by increasing the concentration of $\mathrm{Cl}^-$ ions (and consequently, the amount of $\mathrm{Na}^+$ ions, as required to fulfil the electroneutrality condition at the boundary). By increasing the electrolyte conductivity, the electrolyte potential within the pit will become closer to the reference potential imposed at the top of the domain, as shown in \cref{fig:case6_EI}. This effect by itself does not alter the corrosion rate since the corrosion rate is dictated by the potential difference between the metal and the electrolyte, and not by the potential of the electrolyte itself. However, increasing the electrolyte conductivity not only decreases the potential within the pit, but also slightly increases the potential on the exterior surface, forcing both to attain the same value; the hydrogen and oxygen reactions no longer cause a negative electrolyte potential near the exterior metal surface (with this negative potential slowing the cathodic reaction rates in low conductivity circumstances). As a result, the influence of the reaction products on the electrolyte potential has a lesser effect on the reaction kinetics, slightly accelerating the oxygen reaction and corrosion reactions. Furthermore, since larger concentrations of negatively charged ions ($\mathrm{Cl}^-$ in this case) become available, the pH within the pit is less limited by the electroneutrality condition and ion diffusivities. This, combined with the increase in conductivity and the electric overpotential tending slightly closer to zero within the pit, is also sufficient to increase the corrosion reaction to create an acidic environment within the pit. This increase in hydrogen reaction current further accelerates the corrosion reaction, causing the increase in corrosion rate and thus more negative corrosion current observed when increasing the electrolyte conductivity.

\section{Cathodic reaction support radius and oxygen limits}
\label{sec:Localisation}

While the analysis of the previous section has shown that increasing the radius of the cathodic surface enhances corrosion, this effect is commonly assumed to solely happen up to a limited distance from the pit. To investigate this assumption, simulations are performed using an outer domain radius $r_{\mathrm{domain}}=1\;\mathrm{m}$, while keeping the pit radius at $r_{\mathrm{pit}}=0.5\;\mathrm{cm}$. To further emphasise the role of the support radius of the cathodic reactions, the reference potential and the concentration of ionic species are prescribed at the outer radius instead of the top. In contrast to the previous section, where the boundary conditions approximated a well-mixed reservoir, the boundary conditions used here approximate a thin layer of water on top of a metal surface, with this water layer thickness taken as $h_{\mathrm{domain}}=5\;\mathrm{mm}$. The oxygen boundary condition is still prescribed at the top surface, equal to $C_{\mathrm{O}_2}=0.25\;\mathrm{mol}/\mathrm{m}^3$, and two conditions for this oxygen supply are investigated: (i) an infinite source, and (ii) an initial source lasting for 48 h. This latter finite oxygen condition will highlight the inability of the reactions to become self-sustained in the absence of oxygen. To capture the self-sustainability of corrosion pits more realistically, we add one additional reaction for the metal ions:
\begin{equation}
    \mathrm{FeOH}^{+} + \mathrm{H}_2\mathrm{O} \xrightarrow{k_{feoh}} \mathrm{Fe(OH)}_2 + \mathrm{H}^+
\end{equation}
This uni-directional reaction, using a reaction coefficient $k_{\mathrm{feoh}}=10^{-3}\;\mathrm{s}^{-1}$, will cause each produced $\mathrm{Fe}^{2+}$ ion to eventually reduce to $\mathrm{Fe(OH)}_2$ and create two $\mathrm{H}^+$ ions in the process. These hydrogen ions can then either be used to sustain the corrosion reaction, or be consumed within the water auto-ionisation reaction to retain the local $\mathrm{pH}$. It should be noted that although this reaction produces an insoluble reaction product, this species is not captured within the simulations. Instead, it is assumed that only a limited amount of $\mathrm{Fe(OH)}_2$ is produced, and that this amount does not inhibit the diffusion and surface reactions. The material properties from \cref{tab:properites} are used, together with a pit depth of $h_{\mathrm{pit}}=2\;\mathrm{cm}$ and a pH of 12. The initial and boundary concentration $C_{\mathrm{Cl}^-}$ is varied between $10\;\mathrm{mol}/\mathrm{m}^3$ and $10^4\;\mathrm{mol}/\mathrm{m}^3$. 

\begin{figure}
    \centering
    \begin{subfigure}{8cm}
         \centering
         \includegraphics[width=8cm,trim={0 1.1cm -0.25cm 0},clip]{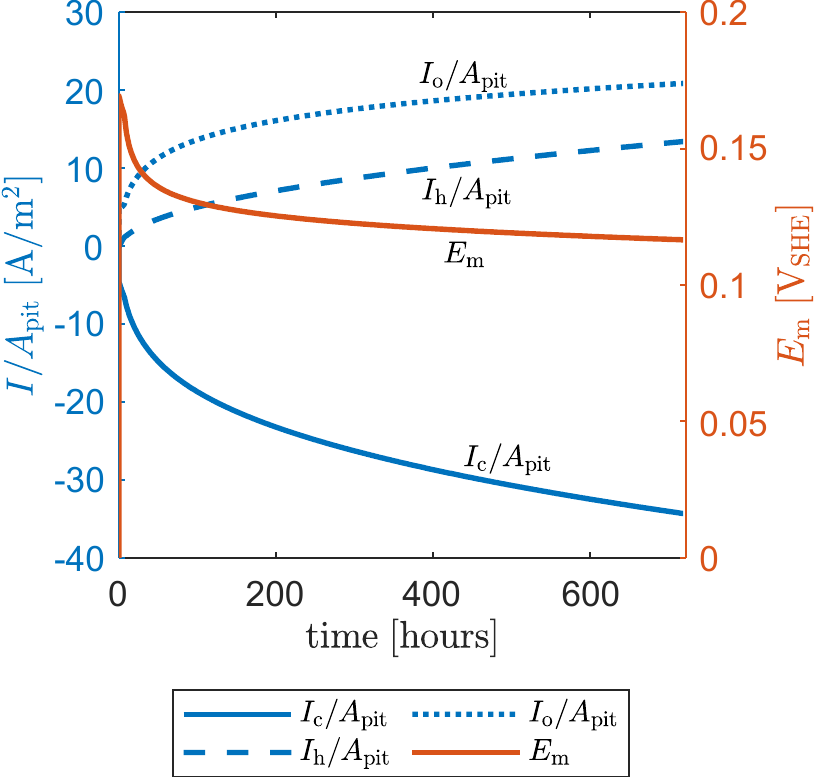}
         \caption{$C_{{\mathrm{Cl}}^-}=10\;\mathrm{mol}/\mathrm{m}^3$}
         \label{fig:sec4_OverTime_10}
    \end{subfigure}
    \begin{subfigure}{8cm}
         \centering
         \includegraphics[width=8cm,trim={-0.25cm 1.1cm 0 0},clip]{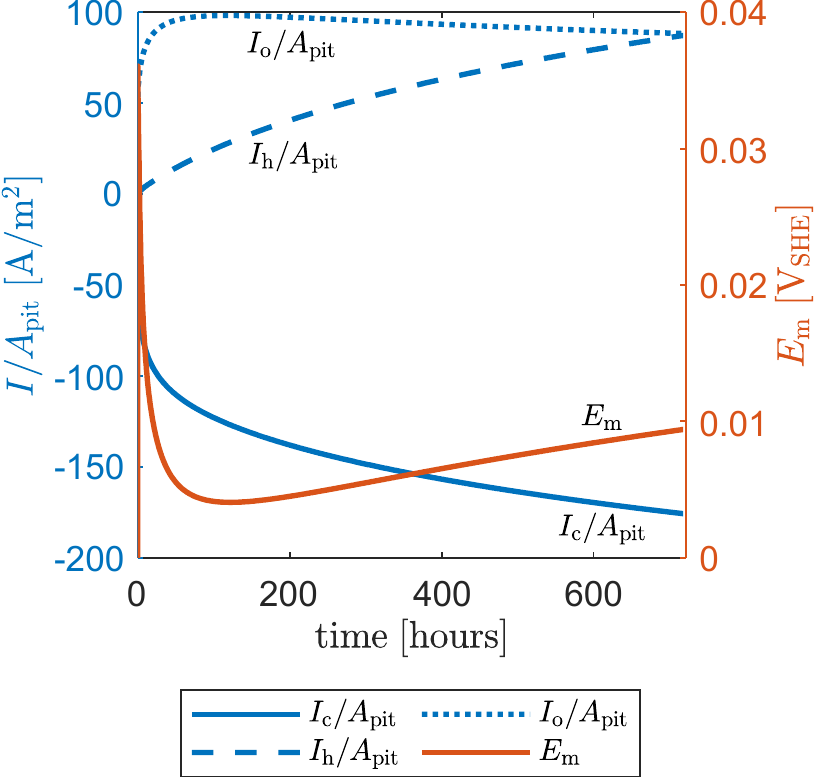}
         \caption{$C_{{\mathrm{Cl}}^-}=1000\;\mathrm{mol}/\mathrm{m}^3$}
         \label{fig:sec4_OverTime_1000}
     \end{subfigure}
    \caption{Evolution of the reaction currents (left) and metal potential (right) over time for $C_{{\mathrm{Cl}}^-}=10\;\mathrm{mol}/\mathrm{m}^3$ (a) and $C_{{\mathrm{Cl}}^-}=1000\;\mathrm{mol}/\mathrm{m}^3$ (b). Unlimited oxygen supply case with $r_{\mathrm{domain}}=1\;\mathrm{m}$.}
    \label{fig:sec4_OverTime}
\end{figure}
\begin{figure}
    \centering
    \begin{subfigure}{8cm}
    \centering
    \includegraphics[width=8cm,trim={0 12cm -3cm 0},clip]{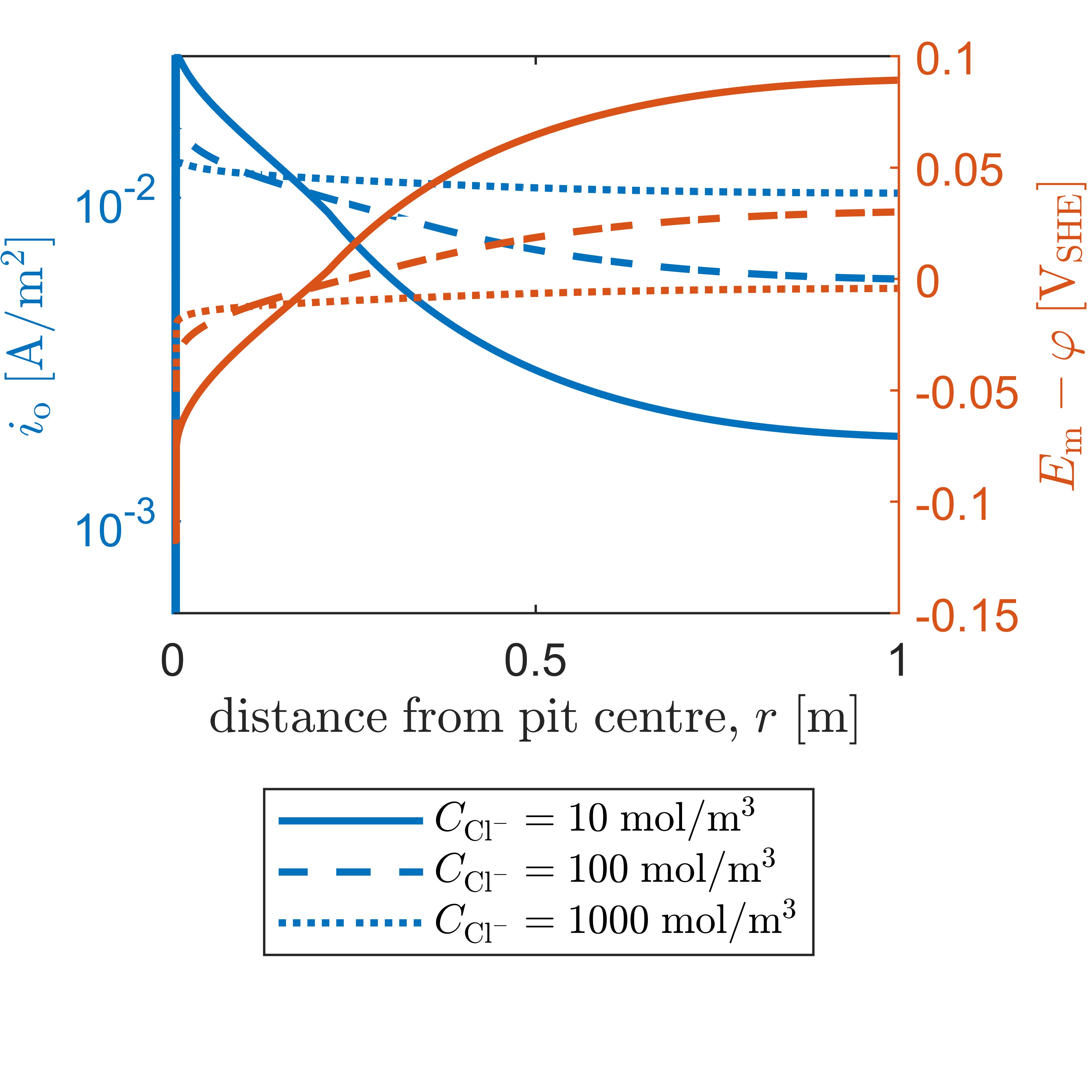}
    \caption{}
    \label{fig:sec4_oxygenRate}
    \end{subfigure}
    \begin{subfigure}{8cm}
    \centering
    \includegraphics[width=8cm,trim={-3cm 12cm 0 0},clip]{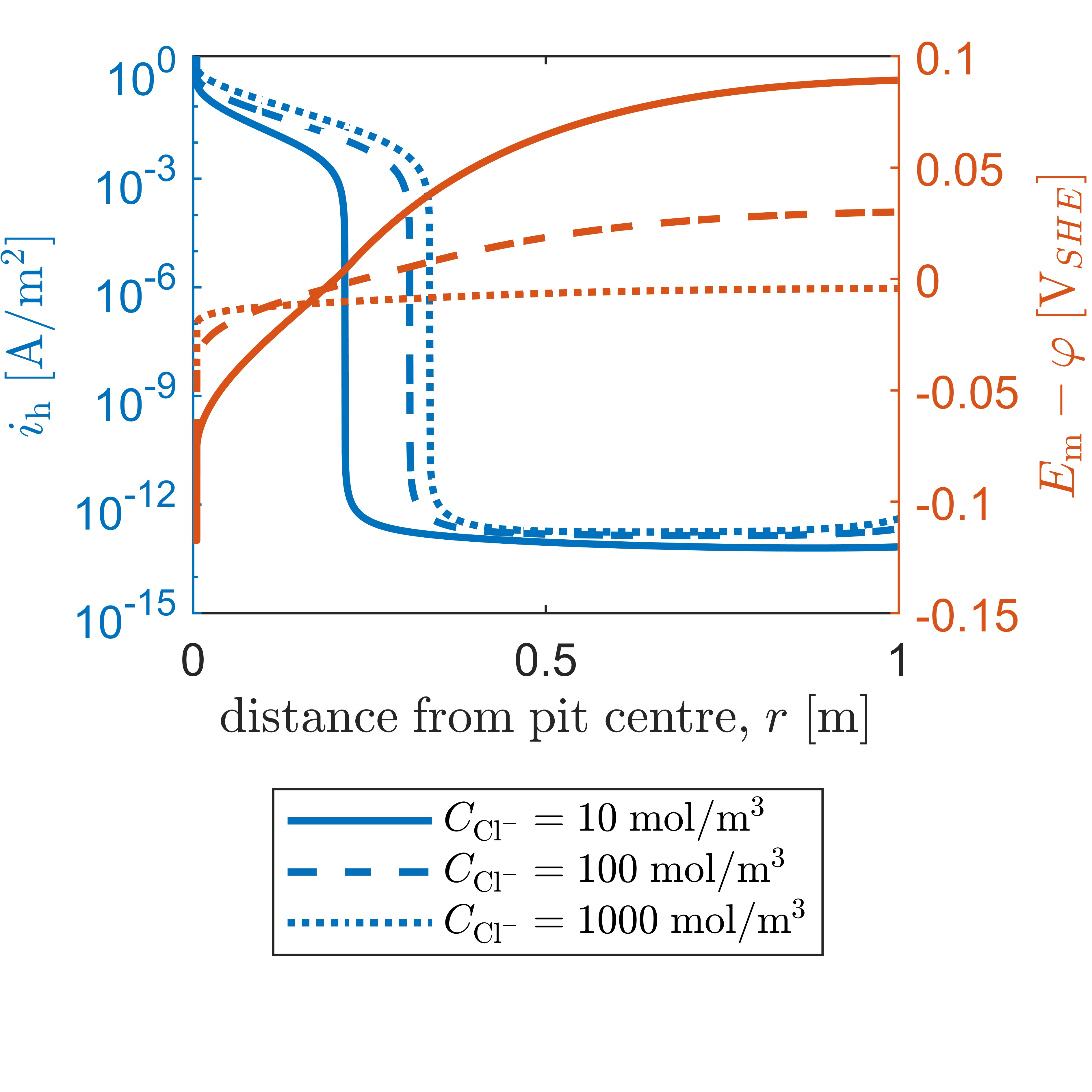}
    \caption{}
    \label{fig:sec4_hydrogenRate}
    \end{subfigure}
    \caption{Spatial distribution of the oxygen (a) and hydrogen (b) reaction currents (blue, left axis), and of the electric potential difference (orange, right axis). Results obtained for an unlimited oxygen supply after $t=30\;\mathrm{days}$.}
    \label{fig:sec4_Unlimitedoxygen}
\end{figure}
\begin{figure}
    \centering
    \begin{subfigure}{8cm}
         \centering
         \includegraphics[width=8cm,trim={0 1.1cm -0.25cm 0},clip]{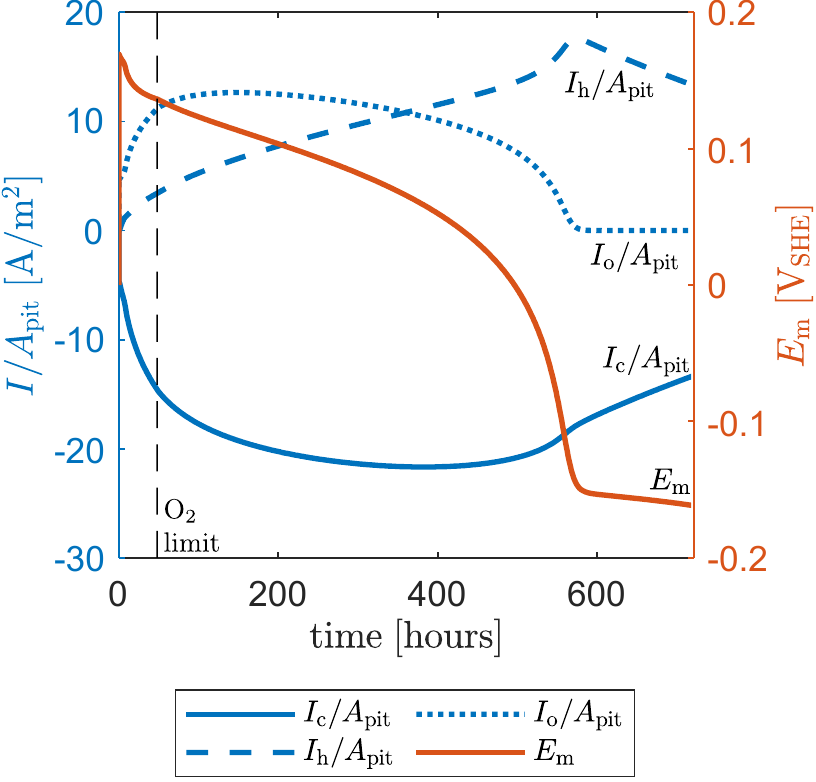}
         \caption{$C_{{\mathrm{Cl}}^-}=10\;\mathrm{mol}/\mathrm{m}^3$}
         \label{fig:sec4_OverTime_limited_10}
    \end{subfigure}
    \begin{subfigure}{8cm}
         \centering
         \includegraphics[width=8cm,trim={-0.25cm 1.1cm 0 0},clip]{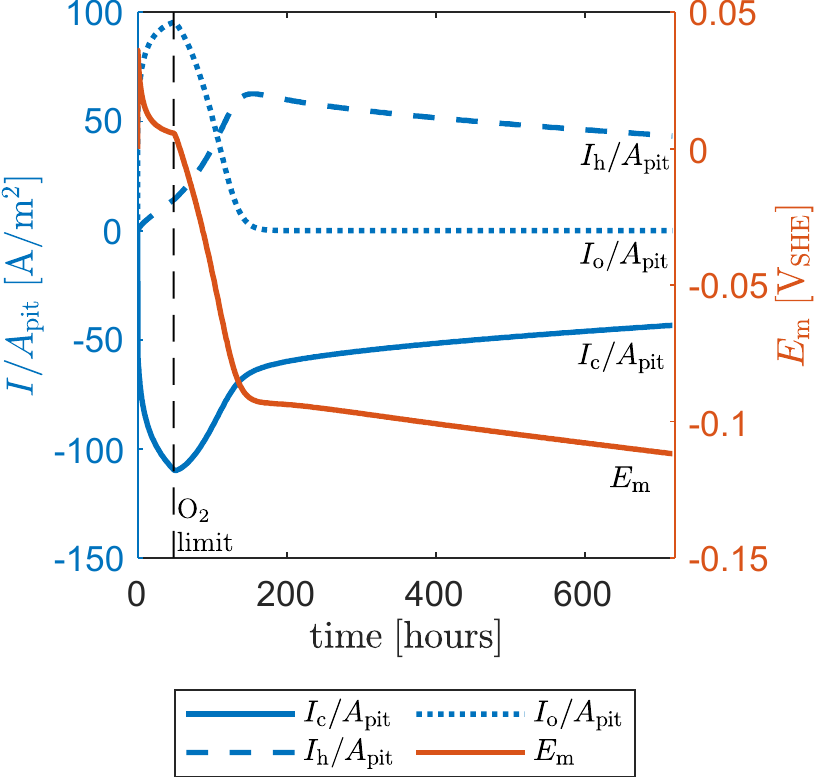}
         \caption{$C_{{\mathrm{Cl}}^-}=1000\;\mathrm{mol}/\mathrm{m}^3$}
         \label{fig:sec4_OverTime_limited_1000}
     \end{subfigure}
    \caption{Evolution of the reaction currents (left) and metal potential (right) over time for $C_{{\mathrm{Cl}}^-}=10\;\mathrm{mol}/\mathrm{m}^3$ (a) and $C_{{\mathrm{Cl}}^-}=1000\;\mathrm{mol}/\mathrm{m}^3$ (b). Here, $r_{\mathrm{domain}}=1\;\mathrm{m}$ and the oxygen is solely entering the domain for the first $48\;\mathrm{hours}$ (up to the black vertical line).}
    \label{fig:sec4_limited_OverTime}
\end{figure}
\begin{figure}
    \centering
    \includegraphics[width=8cm,trim={0 12cm 0 0},clip]{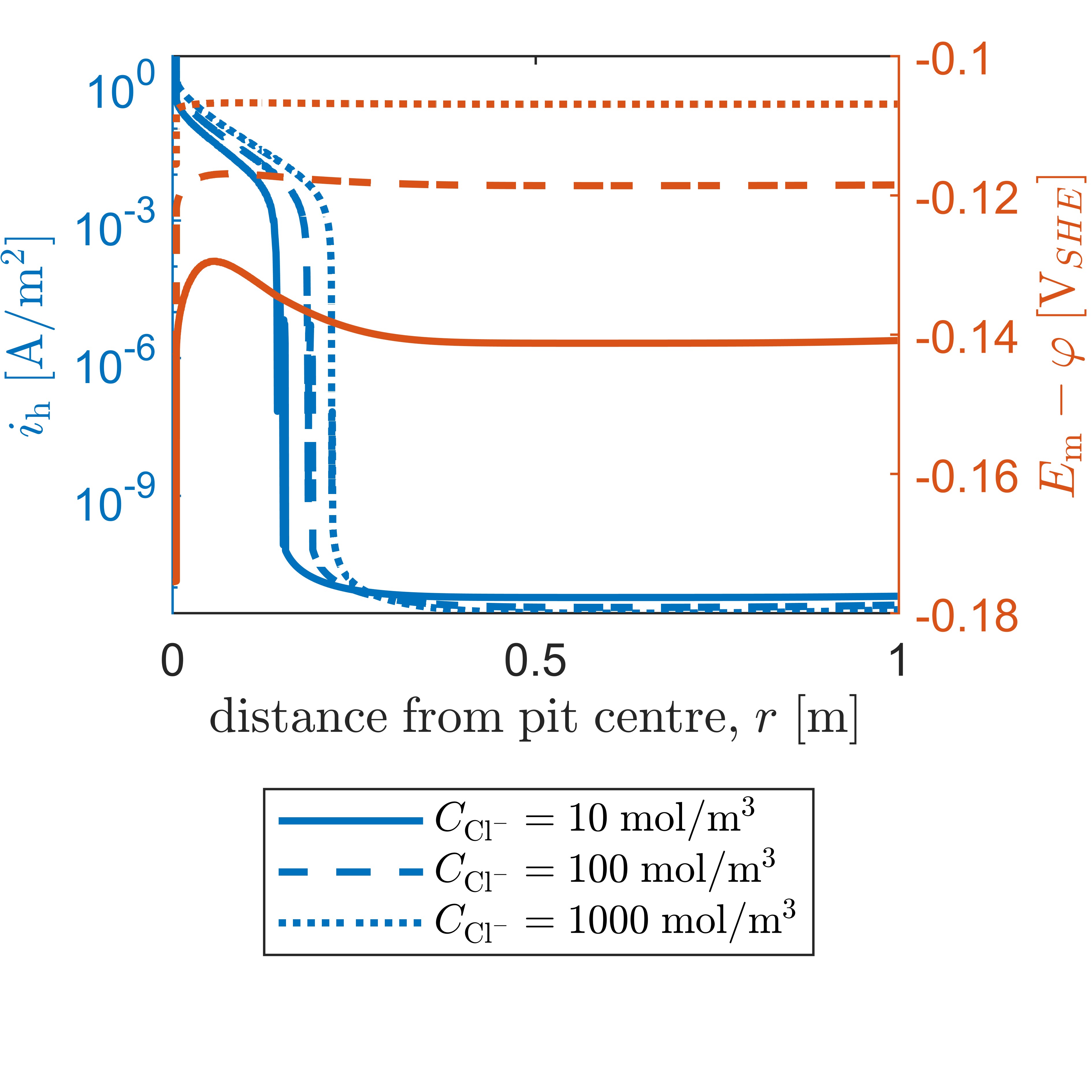}
    \caption{Spatial distribution of the hydrogen reaction currents (blue, left axis) and of the electric potential difference (red, right axis). Results obtained for the $48\;\mathrm{hours}$ oxygen supply case after $t=30\;\mathrm{days}$.}
    \label{fig:sec4_hydrogenRate2}
\end{figure}

The reaction currents computed assuming an infinite oxygen supply are shown in \cref{fig:sec4_OverTime}. When the electrolyte has a low conductivity (due to the low concentration of $\mathrm{Na}^+$ and $\mathrm{Cl}^-$ present, Fig. \ref{fig:sec4_OverTime}a) the reaction currents do not stabilise within 30 days; the oxygen and hydrogen reactions continue to increase (at a similar rate) as the metal potential decreases and the corrosion pit acidifies, which allows for an increased corrosion rate. In contrast, when a higher conductivity is used, Fig. \ref{fig:sec4_OverTime}b, the oxygen reaction stabilises within a day, while the hydrogen reaction rate still increases slowly in time beyond 600 h due to its dependence on the development of the pit environment. As the pit acidifies, favouring the hydrogen reaction, the metal potential increases, which allows for increased corrosion and decreased oxygen reaction rates. The local magnitudes of these reaction rates are shown in Figs. \ref{fig:sec4_oxygenRate} and \ref{fig:sec4_hydrogenRate} for the oxygen and hydrogen reaction rates, respectively. The oxygen reaction occurs throughout the exterior surface. However, for low conductivity conditions ($C_{\text{Cl}^-}=10$ mol/m$^3$, solid line) the potential difference between the electrolyte and the metal is more favourable (negative) near the corrosion pit, causing the oxygen reaction to occur an order of magnitude faster near the pit relative to near the external boundaries. As the conductivity increases, this effect becomes more limited, eventually showing only a limited reduction of the oxygen reaction rate within the 1 m radius of the domain. In contrast, as shown in Fig. \ref{fig:sec4_hydrogenRate}, the hydrogen reaction rate is localised near the pit for all conductivity levels since it strongly depends on the local pH. 

The effects of having a limited supply of oxygen are shown in \cref{fig:sec4_limited_OverTime}. The results show that, for all values of electrolyte conductivity, the oxygen reaction rate eventually becomes negligible. However, the reduced corrosion rate for the $C_\mathrm{Cl}^-=10\;\mathrm{mol}/\mathrm{m}^3$ case (Fig. \ref{fig:sec4_limited_OverTime}a) causes the oxygen to be consumed considerably more slowly, and this causes the changes in corrosion and metal potential to be more gradual. As oxygen is consumed, the metal potential decreases to compensate for the reduced availability of cathodic reactions and this leads to an acceleration of the hydrogen reaction during the duration of the oxygen drop, eventually reaching a maximum when all oxygen is consumed. From this point, the sole cathodic reaction that can balance out the corrosion reaction is the hydrogen reaction. In isolation, this reaction is perfectly balanced with the corrosion reaction and its by-products: The hydrogen reaction consumes two electrons and two $\mathrm{H}^+$ ions, while the corrosion reaction produces two electrons and its by-products create two $\mathrm{H}^+$ ions. However, over time the hydrogen ions within the acidic pit environment diffuse outward, the water auto-ionisation reaction consumes a small amount of $\mathrm{H}^+$, and the equilibrium reactions for the metal ions do not allow the full amount of iron ions to be converted. The combination of these effects causes the pH to increase within the pit, reducing the hydrogen reaction rate, which in turn decreases the metal potential, eventually reducing the corrosion rate. As a result, a slow decline in the corrosion rate is seen once the oxygen becomes unavailable. One additional effect of the lack of oxygen reaction is seen in \cref{fig:sec4_hydrogenRate2}, which shows the local hydrogen reaction rates on the exterior surface. Whereas for the unlimited oxygen study, the complete exterior surface supported the corrosion reaction (through the oxygen and hydrogen reactions), now the sole electron sink required to sustain the corrosion is located within the acidic pit region. As such, all reactions relevant to corrosion are localised in a relatively small area around the corrosion pit, with the majority of the exterior surface not contributing to the corrosion process. 

\section{Conclusions}
\label{sec:conclusions}
In this work, insight has been gained into corrosion behaviour in the absence of external current sources. This has been achieved by presenting a novel computational model capable of handling charge-conservation conditions. This model, numerically implemented using the finite element method, defines the metal potential as an unknown (degree-of-freedom) and includes an additional differential equation to capture the balance between cathodic and anodic currents. Thus, numerical predictions naturally predict how the anodic corrosion current is balanced by the cathodic currents, originating from oxygen and hydrogen reactions. The model is exploited to gain insight into conditions relevant to pitting corrosion by considering the paradigmatic geometry of a pencil electrode test and varying geometric and environmental parameters. The results reveal a strong link between the corrosion rate and the environmental conditions, with the charge conservation causing a direct coupling between the corrosion rate and the cathodic hydrogen and oxygen reaction rates; increasing the oxygen available in the electrolyte, or using a more acidic environment increases these cathodic reaction rates, and in turn causes a significant increase in the hydrogen reaction rate. Key findings of this study include:
\begin{itemize}
    \item The hydrogen reaction rate was seen to have two distinct regions corresponding to an acidic and a basic environment within the pit. For low corrosion rates, the hydrogen produced by corrosion is not sufficient to sustain an acidic environment, and the hydrogen reaction rate is low as a result. Once the corrosion rate passes a threshold, it can sustain an acidic environment, causing a sharp increase in the corrosion rate. 
    \item Using an electrolyte with a higher conductivity due to higher concentrations of non-reacting ions forces the electrolyte potential to be lower within the pit, increasing the overpotential and therefore increasing the corrosion rate.
    \item Changes in geometry can also significantly alter the corrosion rate. Deeper pits result in higher hydrogen reaction currents, slightly increasing the corrosion rate, while a larger height of the exterior domain hinders oxygen transport to the surface, reducing the corrosion rate. Increasing the ratio between the exterior and interior areas further increases the corrosion rate, creating a larger area for the oxygen reaction to occur on, with the electrons consumed in this reaction having to come from a relatively smaller area.
    \item In the cases in which the domain represents a thin layer of water, the cathodic reactions are localised in a limited region around the pit. This region is significantly smaller for lower conductivities.
    \item While the presence of some oxygen is required to initialise the corrosion reaction and the resulting acidic pit environment, this acidic environment is able to sustain the corrosion reaction for some time in the later absence of oxygen. However, eventually the environment will become less acidic, and the corresponding decrease in corrosion rate will not be able to sustain an acidic pit. 
\end{itemize}
Overall, explicitly including the dependence of the corrosion rate on the current conservation of the metal results in a large dependence on the environment, producing significantly different metal potentials compared to those typically imposed during pencil electrode simulations and experiments. These metal potentials in turn result in a significantly different corrosion rate. Since this metal potential is not known a priori, it needs to be either measured or calculated under charge-conservation conditions to accurately obtain the corrosion rates for a prescribed geometry.

\section*{Data availability}
\noindent The \texttt{MATLAB} code used to produce the results presented in this paper, together with documentation detailing the use of this code, are made freely available at \url{www.imperial.ac.uk/mechanics-materials/codes}. Documentation is also provided, along with example files that enable reproduction of the results shown in  \cref{sec:Localisation}. [The code will be uploaded shortly after the paper is accepted]


\section*{Declaration of competing interest}
\noindent The authors declare that they have no known competing financial interests or personal relationships that could have appeared to influence the work reported in this paper.

\section*{Acknowledgments}
\noindent T. Hageman acknowledges financial support through the research fellowship scheme of the Royal Commission for the Exhibition of 1851. E. Mart\'{\i}nez-Pa\~neda acknowledges financial support from UKRI's Future Leaders Fellowship programme [grant MR/V024124/1]. The authors also acknowledge computational resources and support provided by the Imperial College Research Computing Service (http://doi.org/10.14469/hpc/2232).

\appendix
\section{Discretised equations and solution method}
\label{app:disc}
The governing equations are discretised using the finite element method, discretising the concentrations and electrolyte potential through a sum over all elements as:
\begin{equation}
    C_{\pi}=\sum_{el}\mathbf{N}_\mathrm{C} \mathbf{C}_{\pi} \qquad \varphi = \sum_{el} \mathbf{N}_{\varphi} \bm{\upvarphi}
\end{equation}
in terms of their respective degrees of freedom $\mathbf{C}_{\pi}$ and $\bm{\upvarphi}$, and accompanying shape functions $\mathbf{N}_\mathrm{C}$ and $\mathbf{N}_{\varphi}$. Using these shape functions, the Nernst-Planck equation from \cref{eq:massconserv} is cast into its discretised weak form by multiplying with the test function for the concentrations and integrating over the complete domain, resulting in:
\begin{equation} \begin{split}
    \mathbf{f}_{\mathrm{c}\pi} = &\int_{\Omega} \frac{1}{\Delta t}\mathbf{N}_\mathrm{C}^T \mathbf{N}_\mathrm{C} \left(\mathbf{C}_\pi^{t+\Delta t} - \mathbf{C}_\pi^t\right)\;\mathrm{d}\Omega + \int_{\Omega} D_\pi \left(\bm{\nabla}\mathbf{N}_\mathrm{C}\right)^T\bm{\nabla}\mathbf{N}_\mathrm{C}\mathbf{C}_\pi^{t+\Delta t} \; \mathrm{d}\Omega \\
    &+ \int_{\Omega} \frac{D_\pi z_\pi F}{RT} \left(\bm{\nabla}\mathbf{N}_\mathrm{C}\right)^T \left(\mathbf{N}_\mathrm{C} \mathbf{C}_\pi^{t+\Delta t}\right) \bm{\nabla}\mathbf{N}_\varphi \bm{\upvarphi}^{t+\Delta t}\;\mathrm{d}\Omega + \mathbf{R}_{\pi} + \mathbf{\upnu}_{\pi} - \int_{\Gamma} \mathbf{N}_\pi j_\pi \; \mathrm{d}\Gamma = \mathbf{0}
\end{split} \label{eq:f_c} \end{equation} 
\noindent with the flux imposed on the domain boundary defined as $j_\pi$ . Then, the volume reaction rates $\mathbf{R}_{\pi}$ are given by:
\begin{subequations}
\begin{align}
\begin{split}
    \mathbf{R}_{\mathrm{H}^+} &= -\int_{\Omega} k_{\mathrm{eq}}\mathbf{N}_\mathrm{C}^T \left(K_\mathrm{w}-\mathbf{N}_\mathrm{C}\mathbf{C}_{\mathrm{H}^+}^{t+\Delta t}\mathbf{N}_\mathrm{C}\mathbf{C}_{\mathrm{OH}^-}^{t+\Delta t}\right)\;\mathrm{d}\Omega \\&\quad- \int_{\Omega} \mathbf{N}_\mathrm{C}^T \left( k_{\mathrm{fe}}\mathbf{N}_\mathrm{C}\mathbf{C}_{\mathrm{Fe}^{2+}}^{t+\Delta t}-k'_{\mathrm{fe}}\mathbf{N}_\mathrm{C}\mathbf{C}_{\mathrm{FeOH}^+}^{t+\Delta t}\mathbf{N}_\mathrm{C}\mathbf{C}_{\mathrm{H}^+}^{t+\Delta t} \right) \;\mathrm{d}\Omega\end{split} \label{eq:RH} \\
    \mathbf{R}_{\mathrm{OH}^-} &= -\int_{\Omega} k_{\mathrm{eq}} \mathbf{N}_\mathrm{C}^T \left(K_\mathrm{w}-\mathbf{N}_\mathrm{C}\mathbf{C}_{\mathrm{H}^+}^{t+\Delta t}\mathbf{N}_\mathrm{C}\mathbf{C}_{\mathrm{OH}^-}^{t+\Delta t}\right)\;\mathrm{d}\Omega \label{eq:ROH} \\
    \mathbf{R}_{\mathrm{Fe}^{2+}} &= \int_{\Omega} \mathbf{N}_\mathrm{C}^T \left(k_{\mathrm{fe}} \mathbf{N}_\mathrm{C} \mathbf{C}_{\mathrm{Fe}^{2+}} - k_{\mathrm{fe}}' \mathbf{N}_\mathrm{C} \mathbf{C}_{\mathrm{FeOH}^+}^{t+\Delta t}\mathbf{N}_\mathrm{C}\mathbf{C}_{\mathrm{H}^+}^{t+\Delta t}\right) \; \mathrm{d}\Omega \label{eq:RFE} \\
    \mathbf{R}_{\mathrm{FeOH}^+} &= - \int_{\Omega} \mathbf{N}_\mathrm{C}^T \left( k_{\mathrm{fe}}\mathbf{N}_\mathrm{C}\mathbf{C}_{\mathrm{Fe}^{2+}}^{t+\Delta t}-k'_{\mathrm{fe}}\mathbf{N}_\mathrm{C}\mathbf{C}_{\mathrm{FeOH}^+}^{t+\Delta t}\mathbf{N}_\mathrm{C}\mathbf{C}_{\mathrm{H}^+}^{t+\Delta t}  \right)\;\mathrm{d}\Omega \label{eq:RFEOH} \\
    \mathbf{R}_{\mathrm{O}_2}&=\mathbf{R}_{\mathrm{Na}^+} = \mathbf{R}_{\mathrm{Cl}^-} = \mathbf{0} \label{eq:RNACL}
\end{align}
\end{subequations}
and the surface reaction rates are given by:
\begin{subequations}
\begin{align}
    \mathbf{\upnu}_{\mathrm{H}^+} &= \int_{\Gamma_c} \mathbf{N}_{\mathrm{H}^+}^T 2 \nu_\mathrm{h}\;\mathrm{d}\Gamma_c \\
    \mathbf{\upnu}_{\mathrm{OH}^-} &= -\int_{\Gamma_c} \mathbf{N}_{\mathrm{H}^+}^T 4\nu_\mathrm{o}\;\mathrm{d}\Gamma_c\\
    \mathbf{\upnu}_{\mathrm{O}_2} &= \int_{\Gamma_c} \mathbf{N}_{\mathrm{H}^+}^T \nu_\mathrm{o}\;\mathrm{d}\Gamma_c\\
    \mathbf{\upnu}_{\mathrm{Fe}^{2+}} &= \int_{\Gamma_a} \mathbf{N}_{\mathrm{H}^+}^T \nu_\mathrm{c} \;\mathrm{d}\Gamma_a\\
    \mathbf{\upnu}_{\mathrm{Na}^{+}} &= \mathbf{\nu}_{\mathrm{Cl}^{-}} = \mathbf{\nu}_{\mathrm{FeOH}^{+}} =\mathbf{0}
\end{align}
\end{subequations}
with the reaction rates at the surface evaluated from \cref{eq:surf_c,eq:surf_o,eq:surf_h} using the concentrations at the wall and the electric overpotential $\eta = E_\mathrm{m}^{t+\Delta t}-\mathbf{N}_\varphi \bm{\upvarphi}^{t+\Delta t} - E_{\mathrm{eq},\pi}$. All the volume reactions $\mathbf{R}_{\pi}$ and surface reactions $\mathbf{\upnu}_{\pi}$ are integrated over their respective domains and surfaces using lumped integration \citep{Hageman2023}. 

Similarly to the Nernst-Planck equations, multiplying the electroneutrality condition with the test function for the electrolyte potential and integrating over the domain results in:
\begin{equation}
    \bm{f}_\varphi = \int_{\Omega} \sum_\pi z_\pi  \mathbf{N}_\varphi^T\mathbf{N}_\mathrm{C}\mathbf{C}_\pi^{t+\Delta t} \; \mathrm{d}\Omega = \mathbf{0} \label{eq:f_phi}
\end{equation}
The final equation solved for the charge conservation of the metal surface, \cref{eq:current_conservation}, given by an integral over the anodic and cathodic surface areas:
\begin{equation}
    f_{E_\mathrm{m}} = \int_{\Gamma_c} 2F\nu_\mathrm{h} + 4F\nu_\mathrm{o} \mathrm{d}\Gamma_c + \int_{\Gamma_a} 2F\nu_\mathrm{c} \mathrm{d}\Gamma_a= 0 
\end{equation}
Together, these three equations are resolved in a monolithic fashion. Since these equations are nonlinear, they are resolved using a Newton-Raphson scheme, obtaining the concentrations, electrolyte potential, and metal potential increments by solving the following system:
\begin{equation}
    \begin{bmatrix} \frac{\partial \mathbf{f}_{\mathrm{c}\pi}}{\partial \mathbf{C}_{\pi}}  &  \frac{\partial \mathbf{f}_{\mathrm{c}\pi}}{\partial E_\mathrm{m}} & \frac{\partial \mathbf{f}_{\mathrm{c}\pi}}{\partial \mathbf{\upvarphi}} \\ 
    \frac{\partial f_{E_\mathrm{m}}}{\partial \mathbf{C}_{\pi}} & \frac{\partial f_{E_\mathrm{m}}}{\partial E_\mathrm{m}} & \frac{\partial f_{E_\mathrm{m}}}{\partial \mathbf{\upvarphi}} \\ 
    \frac{\partial \mathbf{f}_{\varphi}}{\partial \mathbf{C}_{\pi}} & 0 & 0\end{bmatrix}_i
    \begin{bmatrix} \mathrm{d} \mathbf{C}_{\pi}  \\ \mathrm{d} E_\mathrm{m} \\ \mathrm{d} \bm{\upvarphi} \end{bmatrix}_{i+1} 
    = - \begin{bmatrix} \mathbf{f}_{\mathrm{c}\pi}  \\ f_{E_\mathrm{m}} \\ \mathbf{f}_\varphi \end{bmatrix}_{i}
\end{equation}
This system is iterated until a solution is obtained that satisfies the convergence criteria:
\begin{equation}
     \frac{\epsilon}{\epsilon_0}<10^{-12} \qquad \mathrm{or} \qquad \epsilon_i<10^{-15} \qquad \mathrm{where} \qquad \epsilon_i = \left| \begin{bmatrix} \mathrm{d} \mathbf{C}_{\pi}  & \mathrm{d} E_\mathrm{m} & \mathrm{d} \bm{\upvarphi} \end{bmatrix}_{i+1} \cdot \begin{bmatrix} \mathbf{f}_{c\pi}  & f_{E_\mathrm{m}} & \mathbf{f}_\varphi \end{bmatrix}_{i+1} \right|
\end{equation}
with these convergence criteria chosen strict enough to ensure a well-converged solution for each time increment.

\end{document}